\documentclass[acmsmall]{acmart}

\AtBeginDocument{%
  \providecommand\BibTeX{{%
    \normalfont B\kern-0.5em{\scshape i\kern-0.25em b}\kern-0.8em\TeX}}}

\setcopyright{acmlicensed}
\acmJournal{PACMHCI}
\acmYear{2021} 
\acmVolume{5} 
\acmNumber{CSCW2} 
\acmArticle{427}
\acmMonth{10} \acmPrice{15.00}
\acmDOI{10.1145/3479571}

\usepackage{makecell}
\usepackage{multirow}
\usepackage{caption}
\usepackage{subcaption}
\usepackage{booktabs}
\usepackage{comment}
\usepackage{ragged2e}

\begin{document}

%

\newcommand{\FeedReflect}[0]{NudgeCred}

\title{\FeedReflect{}: Supporting News Credibility Assessment on Social Media Through Nudges}

\author{Md Momen Bhuiyan}
\affiliation{%
  \institution{Virginia Tech}
  \country{USA}
}
\email{momen@vt.edu}
\author{Michael Horning}
\affiliation{%
  \institution{Virginia Tech}
  \country{USA}
}
\email{mhorning@vt.edu}
\author{Sang Won Lee}
\affiliation{%
  \institution{Virginia Tech}
  \country{USA}
}
\email{sangwonlee@vt.edu}

\author{Tanushree Mitra}
\authornote{A portion of this work was conducted while the author was at Virginia Tech.}
\affiliation{%
  \institution{University of Washington}
  \country{USA}
}
\email{tmitra@uw.edu}

\renewcommand{\shortauthors}{Md Momen Bhuiyan et. al.}
\newcommand{\addition}[1]{\textcolor{black}{#1}}
\newcommand{\customquote}[2]{\vspace{3pt}\begin{quote}\small{\textit{``#1''\hspace{5pt}\textbf{(#2)}}}\end{quote}}
\newcommand{\customrq}[2]{\hspace{5pt} {\small{\textbf{#1}. \textit{#2}}}}
\newcommand{\additionnew}[1]{\textcolor{black}{#1}}
\newcommand{\additionred}[1]{\textcolor{black}{#1}}
\newcommand{\additionminor}[1]{\textcolor{black}{#1}}

 \newcommand{\out}[1]{{#1}}
\newcommand{\sang}[1]{\out{{\small\textcolor{blue}{\bf[*** Sang: #1]}}}}
\newcommand{\momen}[1]{\out{{\small\textcolor{green}{[Momen: #1]}}}}



\begin{abstract}
Struggling to curb misinformation, social media platforms are experimenting with design interventions to enhance consumption of credible news on their platforms.
Some of these interventions, such as the use of warning messages, are examples of \emph{nudges}---a choice-preserving technique to steer behavior.
Despite their application, we do not know whether nudges could steer people into making conscious news credibility judgments online and if they do, under what constraints.
To answer, we combine nudge techniques with heuristic based information processing to design \FeedReflect{}--a browser extension for Twitter.
\FeedReflect{} directs users' attention to two design cues: authority of a source and other users' collective opinion on a report by activating three design nudges---\emph{Reliable, Questionable,} and \emph{Unreliable}, each denoting particular levels of credibility for news tweets. 
In a controlled experiment, we found that \FeedReflect{} significantly helped users \additionnew{(n=430)} distinguish news tweets' credibility, unrestricted by three behavioral confounds---political ideology, political cynicism, and media skepticism.
A five-day field deployment with twelve participants revealed that \FeedReflect{} improved their recognition of news items and attention towards all of our nudges, particularly towards \emph{Questionable}.
Among other considerations, participants proposed that designers should incorporate heuristics that users' would trust.
Our work informs nudge-based system design approaches for online media.
\end{abstract}



\maketitle


\section{Introduction}

\begin{figure}
    \centering
    \includegraphics[width=0.90\textwidth]{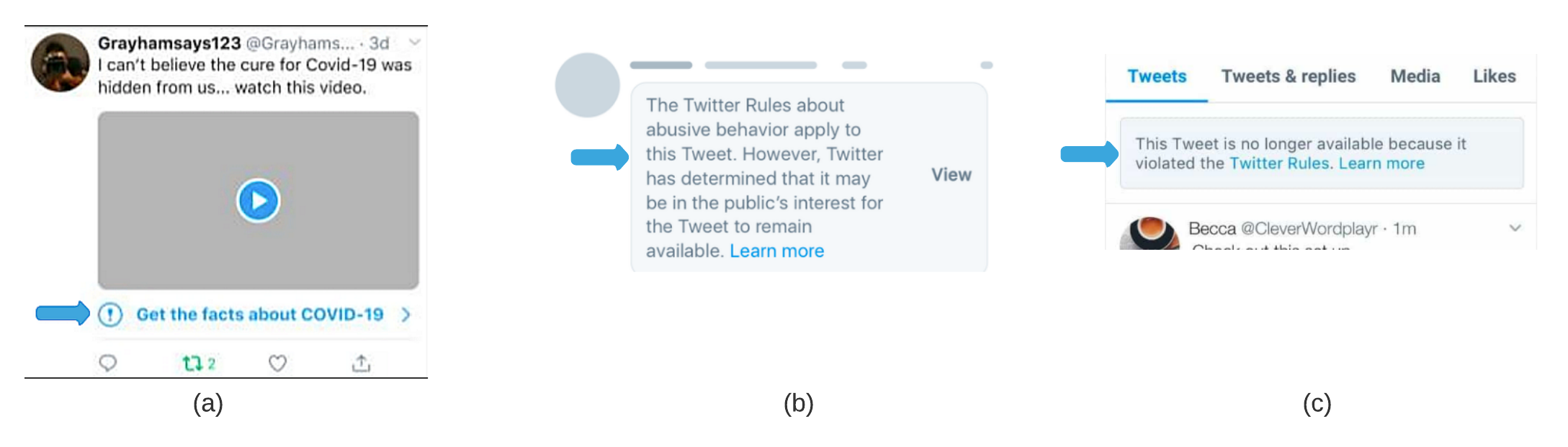}
    \vspace{-15pt}
    \caption{Three types of interventions (marked by blue arrows) currently employed by Twitter to tackle misinformation. Tweet (a) with a link to proper authority regarding COVID-19, (b) with a warning, and (c) removed. Here, both (a) and (b) are examples of nudges. 
    Around the beginning of our work (July 2018), only (c) was operational. Twitter added others later.
    }
    \vspace{-15pt}
    \label{fig:twrules}
\end{figure}

\addition{
Social media platforms have witnessed an unprecedented rise in misinformation around public issues (e.g., COVID-19 and the 2016 US Presidential Election~\cite{Socialme87online,allcott2017social}). 
To tackle, they have responded by experimenting with various design interventions~\cite{Noticeso87online,Updating78online}. 
These include attaching warning labels and links to show additional context from trusted sources (see figure \ref{fig:twrules} for an example).
With an intent to raise awareness and lead users to investigate the veracity of a news item~\cite{Expandin46online}, such design interventions can act as \textit{nudges}---a choice-preserving technique to steer behavior~\cite{sunstein2015ethics}.
\additionnew{Nudges differ from methods such as real-time corrections that often act as mandates and can backfire~\cite{garrett2013promise}.
Furthermore, nudges can overcome the scale issue that many real-time systems face who rely on limited number of expert fact-checkers and quality crowd workers~\cite{medium-crowd}. Despite the benefits,}
there is little empirical evidence whether nudge-based socio-technical interventions affect users' perception of credibility of online information.
Furthermore, what complicates such investigation is that people with strong ideological leaning may resist nudges~\cite{tannenbaum2017misplaced}.
\additionnew{Existing works, again, lack empirical evidence of the effects of ideological leaning on nudges regarding perception of credibility.}
Therefore, considering the constraints under which these nudges may or may not work is crucial. This paper does just that.
}

\begin{figure}[b]
    \centering
    \vspace{-15pt}
    \includegraphics[width=0.99\textwidth]{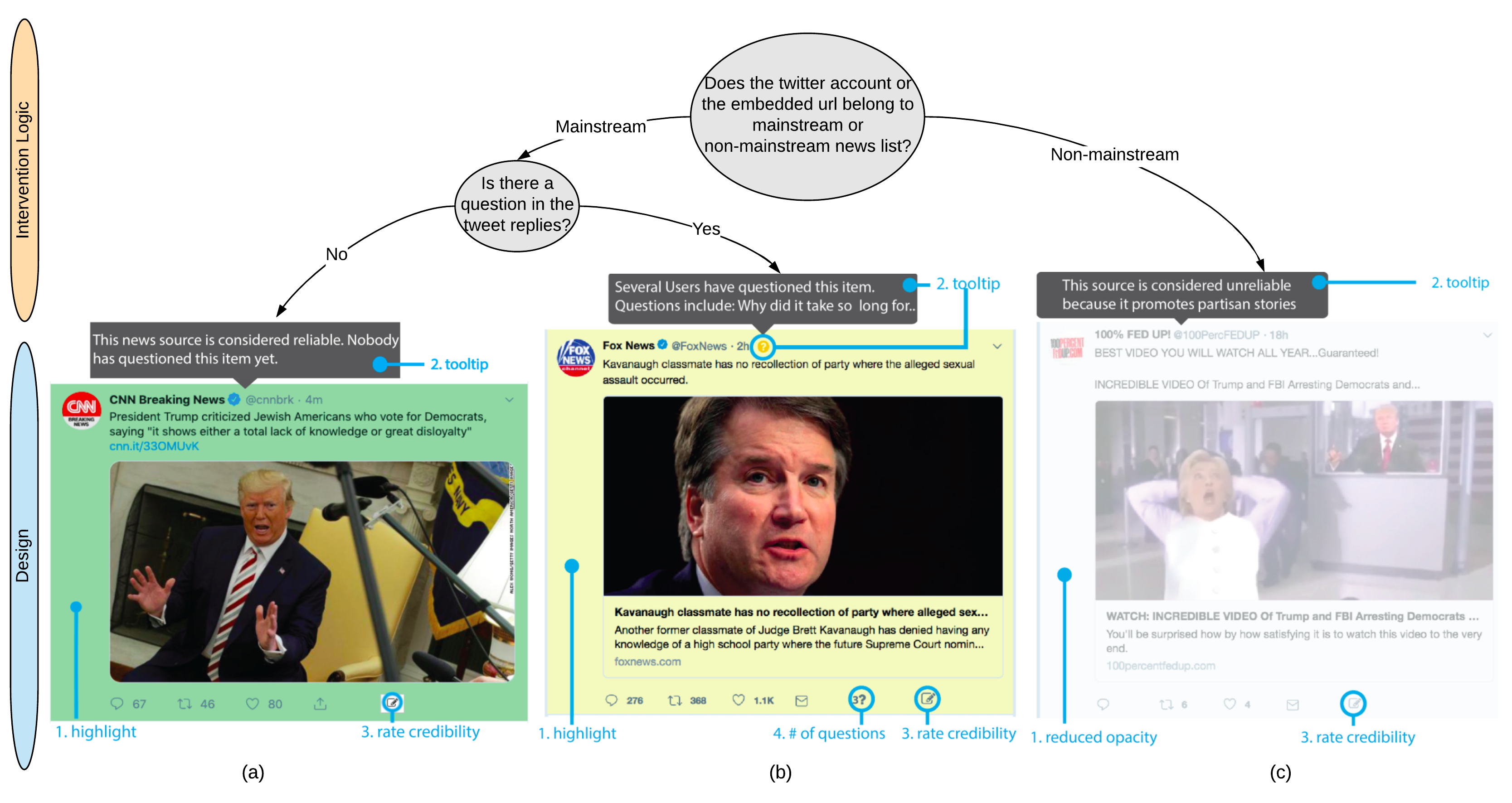}
    \vspace{-15pt}
    \caption{Our nudge design: [Top] A decision tree shows the intervention logic and [Bottom] three nudge designs. (a). The \emph{Reliable} nudge on a tweet from CNN Breaking News without questions in its comment thread. (b). The \emph{Questionable} nudge is applied to a tweet with questions  from Fox News, a mainstream media outlet. (c). The \emph{Unreliable} nudge is activated on a tweet from {\tt 100PercentFedUP.com}, an extremely biased, non-mainstream website. The numbers indicate: (1) a change in background, (2) a tooltip message shown when hovered over, (3) a button to open a survey questionnaire for users to rate the credibility of the news tweet, and (4) a button to show more questions in the comments.}
    \vspace{-15pt}
    \label{fig:em-de}
\end{figure}

For the purpose of our investigation, we design nudges with heuristic cues, i.e., mental shortcuts that people often use to judge credibility~\cite{metzger2010social}.
\additionnew{The choice of heuristic cues over reflective ones reduces cognitive burden on users, given the immense amount of content users see online~\cite{evans2003two}.}
Incorporating design guides from nudge and heuristics literature~\cite{sundar2008main,Todd2000}, we built \FeedReflect{} which operationalizes three design nudges---\emph{Reliable}, \emph{Questionable} and \emph{Unreliable}.
Devised with two heuristic cues---the authority of the source and other users' opinions---each of our nudges designates a particular level of credibility of news content on social media.
\additionnew{These two heuristics comprise both external authoritative sources of information and social interactions of the crowd.
Among the three nudges,}
both \emph{Reliable} and \emph{Questionable} are applied to information originating from mainstream sources on Twitter, while \emph{Unreliable} makes posts from non-mainstream sources less visible (see Figure \ref{fig:em-de}). 
Here, \emph{Questionable} and \emph{Reliable} differentiate between mainstream news items that raised questions in their Twitter replies compared to those that did not. 
Questioned items are highlighted in yellow to warn of the potential controversy in evolving news stories from mainstream media, while those lacking questions are highlighted in green, signifying their reliability.
By directing users' attention to the two heuristics, \FeedReflect{} assists users in making meaningful news credibility evaluations. 

To demonstrate our nudge-based approach, we built \FeedReflect{} as a Chrome extension~\footnote{\additionnew{How \FeedReflect{} works: \url{https://www.dropbox.com/s/2mt4tpdxebccokt/nudgecred\_cropped.mp4}}}. We followed an iterative design process.
We first tested our initial design of \FeedReflect{} by conducting a formative study with \additionnew{16 university students and 36 Amazon Mechanical Turk workers}.
Results from the formative study helped us refine our design and suggested three confounds---political ideology, political cynicism, and media skepticism---that may restrict impacts on users' credibility perceptions.
We then conducted two sets of experiments using our final design: Study 1, a controlled experiment to examine the impact of the nudges with a representative US population ($n=430$); and Study 2, a qualitative field deployment with Twitter users ($n=12$) to gain insight into how we can improve \FeedReflect{}'s design.
Analyzing users' credibility responses from Study 1 revealed that the \emph{Unreliable} nudge significantly reduced users' perceptions of credibility for non-mainstream news sources.
For \emph{Questionable}, users in the treatment group rated news tweets with questions as less credible than the users in the control group, and those without questions as more credible. We did not find any effect of users' political ideology, media skepticism, or political cynicism on the effects of nudges. These outcomes suggest that \FeedReflect{} worked irrespective of these confounds. 
Results from our field deployment (Study 2) show that \FeedReflect{} improved recognition of news content and elicited attention towards all three nudges, particularly \emph{Questionable}. Participants also suggested additional design considerations, such as incorporating heuristics that users would trust, 
applying nudges to share buttons, and using nudges to distinguish news genres and biases. 
To conclude, we offer design directions for news credibility nudging by exploring transparency-mode of thinking nudge categories, other heuristics and nudging methods from prior literature. 
Overall, our contributions include:

\begin{itemize}
    \item A novel approach using heuristic cues to nudge users towards meaningful credibility assessment of news items in social media.
    \item A quantitative evaluation of this approach by examining users' credibility perception while considering three confounds---political ideology, political cynicism, and media skepticism.
    \item A qualitative understanding of the opportunities and challenges of this approach in designing credibility nudges.
\end{itemize}

 
\section{Related Works}
\subsection{Designing for Information Credibility}
At present, social media platforms are taking three approaches to combat misinformation---removing misinformation, reducing their reach, and raising awareness~\cite{RemoveRe27online,Building13online}.
The first line of action falls under the practices of crowdsourced (in-house and community-driven) and technology-assisted moderation~\cite{FacingFa4online,HelpingF52online,Servingh79online} by enforcing established community guidelines~\cite{Introduc14online,Building13online}.
The second approach involves reviews from fact-checking services followed by downranking \cite{HardQues98online,Automati96online} and the application of warning/correction labels~\cite{Addressi62online,Updating78online,Noticeso87online}.
The third approach largely focuses on contextualizing misleading content through design interventions, such as providing source transparency~\cite{HowFaceb12online,Understa80online,Newlabel65online,Designinfeed}, prioritizing content from trusted authorities~\cite{Helpingy12online,AnUpdate67online}, and showing related news stories from various sources~\cite{Replacin94online}.
Some of these interventions also target particular issues (e.g., voting~\cite{Strength26online}) or particular interactions (e.g., message forwarding~\cite{Introduc96online}).
\additionnew{While contextualizing with additional information has its benefits, producing unsorted information only adds to the confusion~\cite{o2002question}. In this regard, we simplify design cues to aid users in distinguishing news coming from mainstream and non-mainstream news sources.}

Aside from these platform-led efforts, researchers have also taken up the challenge of designing tools to aid in assessing information credibility.
These works span several approaches, including fact-checking systems, interventions, media literacy programs and games~\cite{diakopoulos2012finding,caulfield2017web,roozenbeek2019fake,lapowski2018newsguard,maertens2020long}.
There are multiple scholarly efforts for computationally assessing content credibility~\citep{castillo2011information,gupta2014tweetcred,mitra2017parsimonious,qazvinian2011rumor,marinova2020weverify}.
There are some scholarly works on establishing appropriate credibility signals for online content, as well as on designing guides for labeling manipulated media~\cite{zhang2018structured,im2020synthesized,Itmatter54online}.
Some works examine particular crowd-led credibility labeling, including ratings by partisan crowds and the relationship between ratings from crowds and experts~\cite{pennycook2019fighting,becker2019wisdom,bhuiyan2020investigating,mitra2015comparing}. 
\additionnew{Compared to actively seeking labels from the crowd, our work uses both authoritative sources of information as well as passive crowd interaction with content for labelling content credibility in \FeedReflect{}.
This combination helps us overcome the scale issue regarding recruiting quality crowd workers for active labelling.
}

Scholars have employed multiple types of messages as interventions against misinformation, including theory-centered messages~\cite{chen2015deterring}, warning messages~\cite{pennycook2019understanding,bush1994implications}, corrective messages~\cite{pluviano2017misinformation,garrett2013promise,johnson1994sources}, and opposing argument messages~\cite{cook2017neutralizing}. 
Studies examined the efficacy of interventions in various genres of news, including public health~\cite{pluviano2017misinformation,pennycook2020fighting} and politics~\cite{pennycook2019understanding}.
\addition{Some research examined the effectiveness of interventions across countries~\cite{guess2020digital}.}
Others examined effects for interventions across time by offering real-time correction versus delayed retraction~\citep{garrett2013promise,johnson1994sources}. Real-time correction tools utilize various strategies, including mining databases of well-known fact-checking websites (such as, Snopes and PolitiFact) or crowdsourcing fact-checking.
Paynter and colleagues looked into how to strategize corrections by combining several existing techniques (e.g., salience of a graphical element in a warning); they call this approach ``optimized debunking''~\cite{paynter2019evaluation}.
Some suggest that while corrections can be effective, they can also backfire by inadvertently provoking users into attitude-consistent misperceptions~\cite{pennycook2017assessing}. However, others were unable to replicate such backfiring effects~\citep{wood2019elusive}.
Warnings about information credibility have been more successful than corrections and are not prone to the same backfiring effect~\cite{bolsen2015counteracting}.
\addition{While there are various methods (e.g., warnings, reminders, and default options \cite{sunstein2015ethics}) to operationalize nudges, we utilize warnings in \FeedReflect{}.}


 
\subsection{Nudges to Steer Human Behavior}
The concept of \textit{nudges} has been frequently used to steer civic behavior for achieving important societal goals~\citep{halpern2016inside,sunstein2014nudge,sunstein2016people}. 
The idea stems from behavioral economics and psychology, which define it as a ``choice architecture'' that encourages citizens to act in a certain way while allowing them to act in other ways, thereby being a favorable alternative to imposing mandates~\citep{john2009nudge}. 
Such approaches have been highly effective in areas such as environmental protection, financial regulation, and anti-obesity policy~\citep{halpern2016inside,sunstein2014nudge,sunstein2016people}.
In online settings, technology-mediated nudges have been applied for such purposes as encouraging better password management, improving mobile privacy, and encouraging frugal shopping~\cite{bachour2012fast,balebako2011nudging,kankane2018can}.
Comparatively, nudges regarding online news is getting traction recently in works such as Pennycook and colleagues' ``accuracy nudge'' (an accuracy reminder) and Nekmat's ``fact-check alert nudge''~\cite{pennycook2020fighting,nekmat2020nudge}, who investigated impact of nudging on misinformation sharing intention.
\additionnew{This work not only extends existing line of research by employing heuristics to assist credibility judgment, but also shows a method of devising such heuristic cue design.}

\subsection{Heuristic Cues for Credibility}
Cognitive psychologists have long argued that when information overload occurs---as it typically does in online social media environments---humans turn to the cognitively effortless route of peripheral processing~\cite{chaiken1987heuristic,petty1986elaboration}.
\additionnew{While existing HCI works focus extensively on reflective processing, we use automatic or peripheral processing in this study~\cite{adams2015mindless}.}
Peripheral processing means that they depend on heuristic cues, such as the attractiveness of the source or the font color, to evaluate message content~\cite{petty1986elaboration}.
Communication researchers, in response, have offered a well-established list of technology-mediated heuristic cues, highlighted in the MAIN model, which attempts to explain how certain cues influence users' perception of credibility in online contexts~\cite{sundar2008main}.
The MAIN model suggests that four technological affordances influence perceptions of credibility: \textbf{M}odality, \textbf{A}gency, \textbf{I}nteractivity, and \textbf{N}avigability. 
These technological affordances trigger various peripheral cues by which users then judge the credibility of online content.  
For example, the agency affordance focuses on how users perceive source information in computer-mediated contexts. 
Often, the perceived agent or source of authority can be the machine, the user themselves, or the perceived authors of information on a particular website. 
For online news, agency is often attributed to the message's source, and these sources can trigger the \emph{authority} heuristic---the perception that the source is an expert on the subject matter~\cite{sundar2015toward}. 
Similarly, information surrounding a message, such as ratings and recommendations, may also provide contextual information. 
For example, when a group of users likes or shares a news article on social media, the action signals that the group deems the information trustworthy. 
This signal, in turn, can influence users' perception of the information's credibility while serving as a \textit{bandwagon} heuristic~\citep{sundar2008bandwagon,sundar2008main}. 
In summary, the space of all possible heuristics under the four affordances is vast, offering us numerous possibilities for designing credibility nudges.
Among these heuristics, we utilize \emph{authority} and  \textit{bandwagon} heuristics in our nudge design.
We discuss the remaining design possibilities later (see section \ref{sec:design_impl}).

\subsection{Factors Affecting Credibility Perception: Partisanship, Attitude towards Politics, and Media}
Scholars have found numerous factors that may influence information credibility. Based on findings from our formative study (discussed in section \ref{sec:formative}), we contextualize our research on three behavioral confounds---partisanship, attitude towards politics, and attitude towards media.
Historically, scholars have failed to reach a consensus on the role of partisan bias in credibility perception. Earlier work suggested that users perceive unbiased information as more credible compared to one-sided information~\cite{allen1991meta,pechmann1992predicting}. Compared to this result, other research found that users would perceive news conforming to their own attitudes as more credible than unbiased news~\cite{clark1988role,mackie2000impact,metzger2015cognitive}. 
For this reason, users with strong partisan biases might even resist nudges on attitude-challenging news items, rather than be influenced. 
Sunstein hypothesized that a considerably large number of people evaluate nudges based on whether they approve of the underlying political objective, naming this ``partisan nudge bias''~\citep{sunstein2016people,tannenbaum2017misplaced}. 
Hence, we made sure to test our nudge design across a population with balanced partisan affiliations, allowing us to measure the effects of partisanship.

Similar to users' partisan attitude, users' media skepticism can influence their perceptions of the credibility of mainstream and non-mainstream content.
Media skepticism is ``the feeling that the mainstream media are neither credible nor reliable, that journalists do not live by their professional standards, and that the news media get in the way of society rather than help society''~\cite{tsfati2003people}. 
Scholars have found that media skepticism is negatively associated with exposure to mainstream media and positively associated with non-mainstream media~\cite{tsfati2010online}. 
Media is also generally blamed for its role in enhancing institutional distrust by depicting most governmental policies negatively and causing cynicism towards politics~\cite{cappella1997spiral}.
Studies have also demonstrated that users with high media skepticism and political cynicism rated citizen journalists as more credible than mainstream ones~\cite{carr2014cynics}. 
Drawing from these works, 
\additionnew{we examine and provide the first empirical evidence of the effects of political ideology, media skepticism, and political cynicism on credibility nudge.}
 
\section{Formative Study}\label{sec:formative}
We designed \FeedReflect{} in an iterative fashion. Initially, we built a prototype and conducted a pilot study to evaluate it~\footnote{All of our studies have been approved by our Institutional Review Board.}.
\paragraph{Method}
We built our prototype as a Chrome extension with two types of nudges (described in section~\ref{section-designcue}); namely, \emph{Questionable} (tweets highlighted in yellow, which indicate caution) and \emph{Unreliable} (tweets that are less visible).
\additionnew{This extension would alter the Twitter homescreen in real-time when users visit them.
As mentioned in Figure \ref{fig:em-de}, users could click on a survey button added by the extension. Clicking the survey questionnaire button would open a pop-up overlay comprising our study measurements for credibility. We discus them further in section \ref{sec-measurement} (refer Figure \ref{fig:feedreflect_questionnaire} to see how it looked).
}
With this setup, we conducted a study with 52 participants from Amazon Mechanical Turk \additionnew{(n=36) and the university (n=36)} \cite{bhuiyan2018feedreflect}.
\additionnew{For recruitment from the university, we used a university portal available for participant recruitment.
For the MTurk users, we used MTurk portal with some filtering conditions,  such high rate of work acceptance (>95\%), over 18 years of, US resident and familiarity with Twitter.}
In a pre-study survey, we collected users' demographic details, such as gender and political leaning.
Participants were divided into two groups of treatment (seeing tweets with nudges) and control (not seeing any nudge).
In a 2-week study period, we asked our participants to rate the credibility of three to five tweets from their Twitter feeds every day.
\additionnew{We did so by reminding them everyday to spend around 20 minutes on Twitter by completing an MTurk HIT.}
Afterwards, we \additionnew{reached out to 16 users---8 control and 8 treatment users---}to get feedback on our design where 8 of them finally agreed.
\additionnew{In all studies, we compensated our participants adhering to Federal minimum wage requirements~(\$7.25).}

\paragraph{Result}
In our study, we hypothesized that users in the treatment group would rate tweets with both \emph{Questionable} and \emph{Unreliable} nudges as less credible compared to users in the control group.
A Mann-Whitney U test on the credibility ratings  showed that our hypothesis was true for \emph{Unreliable} nudge (avg. cred. (Control) = $0.51$, avg. cred. (Treatment) = $0.43$ and  $Z=210236, p<0.001$, Cohen's $d = 1.291$). 
However, we found the opposite for \emph{Questionable} nudge, i.e., the treatment group rated those tweets as more credible than the control group (avg. cred. (Control) = $0.67$, avg. cred. (Treatment) = $0.71$ and  $Z=502140, p<0.001$, Cohen's $d = 0.188$).
Furthermore, in our post-hoc analyses, we found that for Republican users the effects of nudges were not significant.

To make sense of the discrepancies in our quantitative result, we conducted interviews followed by a thematic analysis.
We identified three themes in the interviews.
First, when asked which news organization users follow, participants showed a trend of interest in ideologically aligned news sources.
While a majority of Democrats mentioned mainstream sources (e.g., CNN, NBC, the New York Times, and the Washington Post), most Republicans named a mixture of mainstream and non-mainstream sources (e.g., the Wall Street Journal, Fox News, Joe Rogan, and Candace Owens).
This trend led us to assume that our intervention may be less effective if it contradicts users' political stances. 
Second, we found several hints that \emph{cynicism towards politics} and \emph{media skepticism} can influence the impact of nudges.
For example, one participant suggested that he prefers news without biases which mainstream media does not do anymore.
Another (Republican) participant expressed frustration that she had to stay away from discussing politics on social media, as she often ran into arguments with others.
If Republicans are indeed more skeptical of mainstream media on the whole, and also equally mistrusting of social media platforms, then our intervention could be perceived as yet another attempt by social media to integrate ideologically motivated interventions into their news feeds. 
Therefore, we decided to examine whether these sentiments of media skepticism and political cynicism adversely affect the interventions. 
Third, consistent with our quantitative result, we found the opposite of the expected reaction to the \emph{Questionable} intervention. For example, a participant responded: \textit{``I found that these tweets [with Questionable intervention] seem ... more accurate than things that I normally read''}.
This conflicting reaction may have stemmed from the lack of a clear hierarchy, i.e., the absence of nudges on more credible news tweets.
Subsequently, we revised our design with a third nudge called \emph{Reliable} (tweets highlighted in green to indicate reliability).
These findings suggest that our initial prototype did not adequately support better news credibility judgments by users, and informed us to consider three confounds (users' political ideologies and attitude towards politics and media) in evaluating our system.

\section{Designing \FeedReflect{}}
\subsection{Design Guides}
To design nudges with heuristic cues, we employ design guides from two strands of literature: the nudge perspective and the heuristic perspective. 

\subsubsection{Nudge Perspective}
To design effective nudges, the literature suggests two primary aspects to consider: the mode of thinking involved (automatic vs. reflective) and the degree of transparency (transparent vs. non-transparent)~\citep{hansen2013nudge}. 

\textbf{Mode of Thinking:} Cognitive psychologists developed \textit{dual process} theories, a set of psychological theories for understanding human decision-making. 
These theories describe two main modes of cognition: \emph{automatic} and \emph{reflective}~\citep{evans2003two}. 
The automatic mode is fast and instinctive. It uses prior knowledge or past repeated behavior and minimal cognitive capacity to decide on actions. 
Reflective thinking, on the other hand, is slow and effortful. It uses greater cognitive capacity to make a goal-oriented choice by critically examining the effects of choices before selection. 
 
\textbf{Transparency:} Scholars introduced epistemic transparency (i.e., whether users would understand the purpose of a nudge) to divide existing nudge designs into two categories: transparent and non-transparent~\citep{hansen2013nudge}. Thaler and Sunstein adopted transparency as a guiding principle for nudges~\citep{thaler2008nudge}. This is because of the concern that a designer may manipulate people into their own preferred direction using systems for behavioral changes. 

Using the combination of these two dimensions, Hansen and Jespersen grouped existing nudges into four categories: reflective transparent, automatic transparent, reflective non-transparent, and automatic non-transparent~\cite{hansen2013nudge}. 
In designing technology-mediated nudges for credibility, we pick one quadrant from these categories: \emph{transparent} nudges with the \emph{automatic} mode of thinking. 
We chose the automatic mode as it requires less cognitive effort to process information, especially given the information overload in social media and the instant nature of media consumption.
\additionnew{Scholars in the past argued that use of automatic mode over reflective mode for design could address two potential problems--lack of motivation and lack of ability--that typically restrain users from performing tasks such as critically evaluating credibility~\cite{adams2015mindless}.
Furthermore, our design does not prevent users from critically reflecting on the news content.}
We chose the \emph{transparent} design to explicitly reveal the motives behind it. 
We later discuss the potential for nudge designs in the remaining three quadrants (see section \ref{sec:design_impl}). 

\subsubsection{Heuristic Perspective}
This work applies heuristics to design nudges for social media in order to enhance users' perceptions of the credibility of news. 
Cognitive psychologists have proposed models of how effective heuristics work~\cite{Koehler2004}. One of the models, called \textit{Fast and Frugal Heuristics} suggests that users should be able to make inferences using ``fast, frugal, and accurate'' heuristics when faced with environmental challenges (e.g., information overload)~\cite{Todd2000}. According to Todd et. al., simple heuristics work when they follow two principles: they exploit the structure of the environment and are robust. In social media, structures include sources of news items, popularity (indicated by the number of shares or replies), and the way that information is organized by time and personal interactions. 
They argued that heuristics that exploit existing structured information can be ``accurate without being complex''~\cite{Todd2000}. 
Another success criteria for heuristic design is the robustness of the decision model. A computational strategy utilizing a limited set of information can yield more robustness~\cite{Todd2000}. Employing these principles, our design includes only two heuristics, outlined below. These heuristics seem useful to users to investigate misinformation~\cite{geeng2020fake}.

\begin{itemize}
    \item \textit{Authority Heuristic}: We limit the source of news to a handful of known organizations followed by a binary classification of the organizations. 
    \item \textit{Bandwagon Heuristic}: We utilize the conversational structure (or replies) of the environment as an indicator of credibility assuming a skew in the reply distribution. 
\end{itemize}

\subsection{Outlining the Design}\label{section-designcue}
Our design of \FeedReflect{} is built on the idea of applying subtle heuristic cues in certain contexts in social media. 
It is powered by three types of socio-technical interventions---\emph{Unreliable}, \emph{Questionable}, and \emph{Reliable}. 
Using the principles of fast and frugal heuristic design, our design uses a two-level decision tree with two heuristics (see figure \ref{fig:em-de}). 
They are triggered based on whether a news tweet originates from an official authority. 
Thus, the first step of our tool design relies on the \emph{authority heuristic}. 
Communication scholars have long argued that revealing the official authority of content results in applying the authority heuristic in credibility judgments~\cite{sundar2008main}.
We apply the authority heuristic by differentiating between mainstream and non-mainstream news tweets.
We do not apply nudges to tweets that do not come from mainstream and non-mainstream sources.
\addition{We opt to use source-based credibility annotation due to the challenging nature of annotating article-level credibility.
While we may flag some accurate articles from non-mainstream media in this method, other work demonstrated that this number could be few (14\%) compared to the accuracy (82\%)~\cite{shao2018spread}. For such false-positives, users still have the opportunity to fact-check them.}

\addition{To flag inaccurate content from mainstream media, we apply another criteria of whether someone replied to a mainstream tweet with a question.}
Literature suggests that such questions, depending on users' prior knowledge of the subject matter, can instigate curiosity and motivate them to investigate to a varying degree~\cite{litman2005epistemic}. 
We employ this property by showing the number of questions as a \emph{bandwagon heuristic} (``if others question this story, then I should doubt it, too'') on mainstream news tweets. Thus, our study had three nudges: mainstream news tweets that did not have questions raised about their content (\emph{Reliable}), mainstream news tweets that had questions raised about their content (\emph{Questionable}), and non-mainstream news tweets (\emph{Unreliable}). 
To employ epistemic transparency in the design, our design includes a tooltip with the reasoning behind each nudge. Figure \ref{fig:em-de} shows the overall design. Before delving into each intervention, we propose a classification of news sources that enables our application of the \emph{authority heuristic}. 

\subsubsection{Classifying Authority of News}
Our nudge-based interventions work on two types of news sources: mainstream and non-mainstream. 
In journalism and communication literature, the term ``mainstream media'' lacks an official definition. 
For the purposes of our study, \textbf{mainstream news sources} are sources recognized as more reliable in prior scholarly work. 
Opting for a heuristic approach, we use such existing literature to create a reliability measure which may later be replaced; this is not our primary contribution.
In this approach, the first two authors iteratively collected a list of mainstream news websites by referring to two prior works, including Pew Survey and NPR~\citep{mitchell2014political,bigsix}. 
Next, we refined our list with the help of our in-house journalism and communication media expert by referring to the most circulated and the most trusted news sources~\citep{Top10USD,mitchell2014political}, subsequently removing a news aggregator (Google News) and a local source (AMNewYork). 
Table \ref{tb:mainstream_list} shows a sample. 
Every source on our final list of 25 news sources follows standard journalistic practices in news reporting~\citep{apnews}.

\begin{table}[t]
\vspace{-7pt}
\sffamily \scriptsize 
\parbox{.3\linewidth}{
\centering
\begin{tabular}{r c c l}
		 \\
		\\
		\multicolumn{3}{c}{\additionminor{Mainstream Source}} \\ 
		\midrule
The Economist       & CNN   & The Blaze       \\
New York Times  & NPR   & BBC    \\
Washington Post & MSNBC         & Fox News\\
Chicago Tribune & WSJ & Politico   \\
New York Post  & Newsday & NY Daily    \\

		\hline
	\end{tabular}
    \caption{Example sources in our \emph{mainstream news} category. 
    }
    \label{tb:mainstream_list}
}
\hfill
\parbox{.61\linewidth}{
\centering
\begin{tabular}{lll}
		Website & Category & Inaccuracy Type Message \\
		\midrule
		abcnews.com.co & Fake news  & misinformation\\
		breitbart.com & Extreme Bias & partisan stories\\
		americantoday.news & Rumor Mills & rumor \\
		infowars.com & Conspiracy Theory & conspiracy\\
		rt.com & State News & state propaganda\\
        \hline
	\end{tabular}
	\caption{Example \emph{non-mainstream news sources} and their categories of reporting inaccuracy. The tooltip messages read: ``This source is considered unreliable because it promotes $<$InaccuracyType$>$''.}
 	\label{tb:nonmainstream_types}  
}
\vspace{-24pt}
\end{table}

For \textbf{non-mainstream news sources}, we refer to {\tt Opensources.co}, a professionally curated list of websites known to spread questionable content~\cite{OpenSour}.
Each source in this list is categorized based on its level of information reliability (e.g., `extreme bias,' `rumor,' `clickbait'). 
The curators manually analyzed each source's domain-level characteristics, reporting and writing styles before assigning a particular category. 
From this list, we remove the `politics' and `reliable' categories to retain sources which were explicitly labeled as promoting unreliable news, a total of 397 sources spanning 10 categories. 
Table \ref{tb:nonmainstream_types} shows a sample from this list. 
We do not intervene in the rest of the sources that do not fall into these two categories.
Using this notion of mainstream and non-mainstream news sources, we apply three nudges.

\subsubsection{Three Nudges} 
The \emph{\textbf{Unreliable}} nudge detects whether a tweet comes from an unreliable authority. Our design applies this nudge by examining whether a tweet from a user's feed originates from a non-mainstream news site and subsequently reduces the item's opacity, rendering it harder to read. We call these tweets unreliable non-mainstream tweets ($T_U$) [See (c) in figure \ref{fig:em-de}]. To instigate epistemic transparency, \emph{Unreliable} provides an explanation of its action through a tooltip message: ``This source is considered unreliable because it promotes $<$InaccuracyType$>$.'' Table \ref{tb:nonmainstream_types} shows the list of $<$InaccuracyType$>$ messages based on the source and its category in \textit{opensources.co}. 

The \emph{\textbf{Questionable}} nudge is applied to mainstream news tweets ($T_Q$) when at least one question is raised about the information in the corresponding Twitter reply thread\footnote{Though Twitter has recently rolled out a threaded reply structure, at the time of the study, it did not exist. Thus, we only took direct replies into account.}. 
Prior studies suggest that less credible reports on an event are marked by questions and inquiries~\citep{zhao2015enquiring,mitra2017parsimonious}.
\addition{To detect questions, our algorithm is kept intentionally simple. Our algorithm looks for ``?'' mark to identify questions---a simple but transparent method that is understandable by users. 
It is worth noting that our focus is not to develop the most sophisticated algorithm to detect questionable news, rather testing the effectiveness of nudge in credibility assessment.
In that regard, using the simple heuristic serves its role and have benefits in simplicity and transparency for users to understand. 
While investigating advanced natural language parsing methods to identify relevance of the questions to the news article or more advanced machine learning techniques to detect questions is worth looking into~\cite{lee2017end,xiong2016dynamic}, such investigations would require significant work, perhaps amounting to a separate full contribution. Hence we leave those as future paths to pursue.  
Instead our approach works as a minimum baseline to identify questions.
}
To make users aware of these questioned mainstream tweets, the \emph{Questionable} nudge is applied by changing the background color of the tweet to yellow while showing the number of questions (see (b) in figure \ref{fig:em-de}). 
By showing this number, we promote a collective endorsement that multiple users have doubts about this news~\cite{sundar2008bandwagon}. 
Additionally, a tooltip message offers transparency by explaining the reason behind the nudge activation.
For $T_Q$, the tooltip message follows the format: ``Several users have questioned this item.
Questions include: $<$first reply tweet with a question$>$'' (e.g., for a tweet containing a news report with missing details, such as time of an event, a reader may have replied to ask: ``When did this happen?''), thus directing further attention to other users' comments in an effort to stimulate the \textit{bandwagon heuristic}.
The bandwagon effect has been demonstrated to be powerful in influencing credibility judgments~\citep{knobloch2005impact}.


The \emph{\textbf{Reliable}} nudge is triggered when the source of the news tweet is an official, mainstream source and was not questioned in the replies; specifically, reliable mainstream tweets ($T_R$) were emphasized with a green background highlight (see figure \ref{fig:em-de} (a)).
A tooltip message is formatted for $T_R$ as follows: ``This tweet seems more reliable. Nobody has questioned this item yet.'' 
The colored highlights and the corresponding tooltip messages create contrast within the mainstream news tweets, helping users navigate them better.
Note that we included this nudge in response to the findings from the formative study.

\section{Study 1: Evaluating Impact on Perceptions of Credibility in a Controlled Setting}\label{sec-study1}
To evaluate our design, we conducted two studies. Study 1 evaluates  impact on credibility perception in a controlled setting while Study 2 is a field deployment.
In Study 1, we examine three research questions on the effect of nudges on users' credibility perceptions in a controlled setting simulating a Twitter feed with multiple tweets for each nudge.

\vspace{5pt}
\customrq{RQ1}{Can heuristic-based design nudges on an online social news feed help users distinguish between reliable, questionable, and unreliable information?}

\customrq{RQ2}{Do users' partisan attitudes affect their responses to credibility nudges?}

\customrq{RQ3}{Do users' political cynicism and media skepticism affect their responses to credibility nudges?}


\subsection{Method (Study 1)}
\begin{table}[t]
\begin{center}
	\centering
	\sffamily \scriptsize 
	\vspace{-8pt}
	\begin{tabular}{rrlcrrl}
		Source Type & Twitter Account &  Political Bias & & Source Type &  Twitter Account & Political Bias  \\
		\cline{1-3} \cline{5-7}
		Mainstream & CNN Breaking News & Left &  & Non-mainstream & Daily Kos & Left\\
		Mainstream & NY Post & Right & & Non-mainstream & Breitbart & Right\\
		Mainstream & Politico & Center & & Non-mainstream & Zero Hedge & Conspiracy\\
        \cline{1-3} \cline{5-7}
	\end{tabular}
 	\caption{Example news sources and their political biases.}
 	\vspace{-22pt}
 	\label{tb:pol_bias}  
\end{center}
\end{table}
\subsubsection{Selecting news tweets for a controlled environment} 
For this study, we simulated a Twitter feed with a fixed set of tweets which would be shown to every user. To simulate a realistic Twitter feed, we selected news sources from our previously compiled list of mainstream and non-mainstream sources, and then selected several tweets from each source (see Table \ref{tb:mainstream_list} and \ref{tb:nonmainstream_types}). 
We used a balanced approach to select sources with left-wing, centrist, and right-wing biases.
For bias categorization, we used {\tt mediabiasfactcheck.com}, a source used in other scholarly works~\cite{AboutMed14online,kiesel2019semeval}.
For each news source under each bias category, we first found the source's Twitter handle. We retained the Twitter accounts that had the greatest numbers of followers\footnote{We excluded Fox News' Twitter handle due to its inactivity for several months until the time of the study.} (a mark of popularity of the news source on Twitter).  
Table \ref{tb:pol_bias} shows sample Twitter accounts with their perceived political biases.
For each source, we selected the tweet within the last 48 hours that had the highest number of shares.
With three nudges working across three political leaning categories, our feed comprised 9 tweets (3 political leanings $\times$ 3 nudges). 
\additionnew{Appendix \ref{sec-example} shows several tweets from this list.}
To add variation, we created another set of 9 tweets using the second most followed Twitter accounts from our list of news sources, resulting in a second feed for our controlled environment.
Users were randomly shown one of the two feeds, totaling 9 tweets in each case.
To evaluate our RQs, we needed to measure users' credibility perception, political ideology, political cynicism, and media skepticism.

\begin{figure}
    \centering
    \includegraphics[width=.95\linewidth]{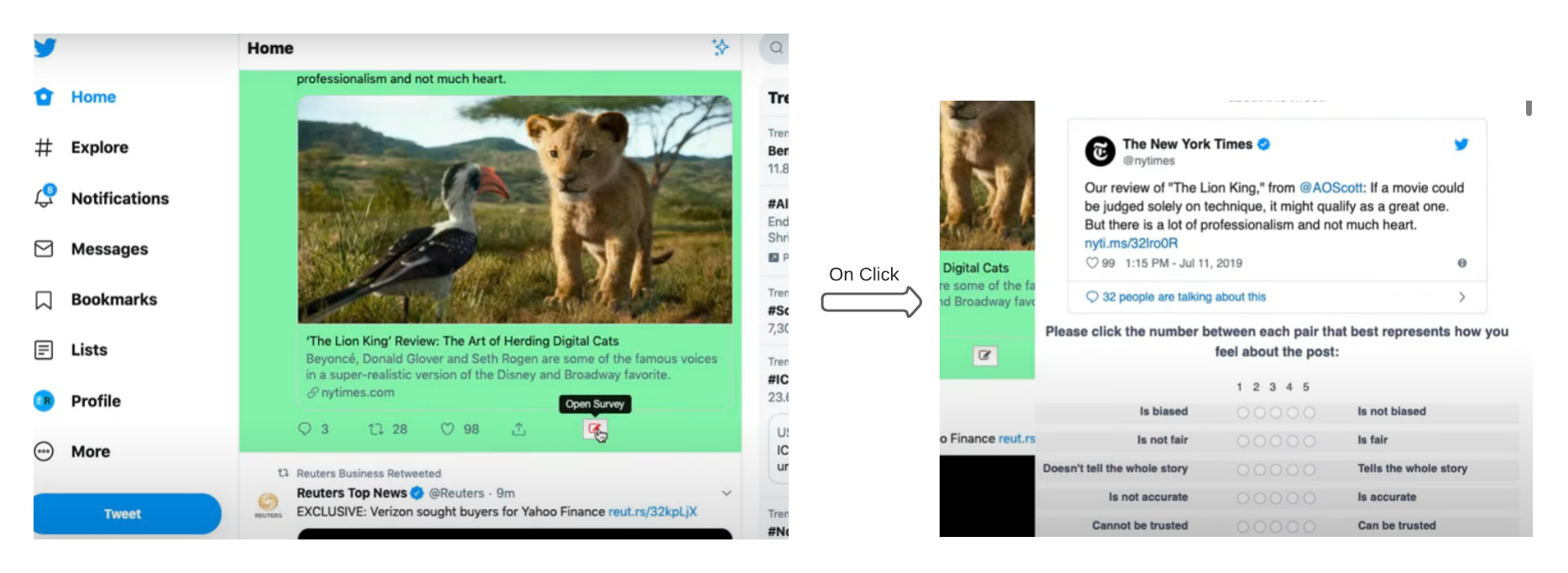}
    \caption{\additionnew{Screenshot of how clicking on the survey button would pop open the five-item credibility questionnaire.}}
    \label{fig:feedreflect_questionnaire}
\end{figure}

\begin{figure}[t]
\begin{minipage}{\textwidth}
  \begin{minipage}{0.40\textwidth}\vspace{0pt}
    \centering
    \sffamily \scriptsize 
    \begin{tabular}{rp{0.15\textwidth}}
    \textbf{\additionnew{Item}} & \textbf{\additionnew{IRR}}\\
    \midrule
    \additionnew{Is/not biased} & \additionnew{0.83}\\
    \additionnew{Is/not fair} & \additionnew{0.79}\\
    \additionnew{Does/not biased tell the whole story} & \additionnew{0.80}\\
    \additionnew{Is/not accurate} & \additionnew{0.79}\\
    \additionnew{Can/not trusted} & \additionnew{0.79}\\
    \hline
    \end{tabular}
    \captionof{table}{\additionnew{IRR of the five-item questionnaire on credibility in the formative study.}}\label{tb:survey}
  \end{minipage} 
  \hfill
  \begin{minipage}{0.58\textwidth}\vspace{0pt}
    \centering
    \sffamily \scriptsize 
    \begin{tabular}{rp{0.85\textwidth}}
    \midrule
     \textbf{Pol.}& 1. Elected officials put their own interests ahead of public's interest\\
    \textbf{Cyn.} & 2. It seems like politicians only care about special interests \\
    \hline
    & 1. The media provide accurate information \\
    \textbf{Med.} & 2. The media provide trustworthy information \\
    \textbf{Skep.}& 3. The media deal fairly with all sides \\
    & 4. The information provided by the media needs to be confirmed \\   
    \hline
    \end{tabular}
      \captionof{table}{Items used in measuring political cynicism and media skepticism. We used a five-point Likert scale (Strongly Agree -- Strongly Disagree) with a ``Don't know'' option.}
    \label{tab:medpol}
    \end{minipage}
  \end{minipage}
\vspace{-22pt}
\end{figure}
\subsubsection{Measuring News Credibility, Political Ideology, Political Cynicism \& Media Skepticism}\label{sec-measurement}
%
We used a five-item questionnaire by Meyer et. al.~\cite{meyer1988defining} to measure users' perceptions of credibility for every news tweet (see Figure \ref{fig:feedreflect_questionnaire}). 
In our formative study, we found this measure had a high Cronbach $\alpha$ ($\alpha = 0.95$) and individual inter-item correlations (see Table \ref{tb:survey}), showing a high level of internal consistency.
To capture partisan attitudes, we survey participants for their political ideology on a seven-point Likert scale ranging from ``strong Republican'' to ``strong Democrat''. 
We survey participants on media skepticism and political cynicism using a validated questionnaire from journalism scholarship (see Table \ref{tab:medpol})~\citep{carr2014cynics}. For both variables, we average the responses across questions and use the median to create a binary response variable with values ``low'' and ``high.'' Note that we had to revert the first three media skepticism questions before averaging.
\begin{figure}[b]
\centering
\includegraphics[width=0.90\textwidth]{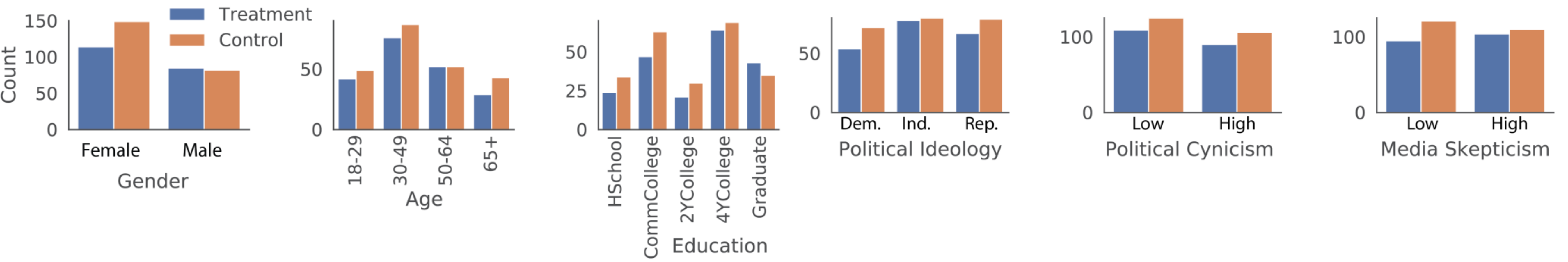}
\vspace{-12pt}
\caption{Distribution of demographics, political ideology, political cynicism, and media skepticism in our participants in Study 1.}
\vspace{-15pt}
\label{fig:demo_all}
\end{figure}
\subsubsection{Recruitment}
Our study participants were recruited starting the third week of July 2019 and spanning a period of three weeks. 
We required three qualifications for user participation: (1) age of 18 or older, (2) US resident, and (3) familiarity with Twitter.
\addition{This choice of US population was purposeful due to the difficulty in measuring our confounds across global population. Users’ political leaning has different meanings in different countries (e.g., political left-right are different in the US and Europe). Similarly, levels of skepticism/cynicism might vary by country. We focused on US-population due to the availability of well-established measurements for our confounds from the communication literature \cite{carr2014cynics,meyer1988defining}.}
We recruited 430 users from \textit{Qualtrics}, well-balanced by partisan affiliations.
Figure \ref{fig:demo_all} shows their demographics.
This sample is mostly balanced across gender, age, and education, with a slight skew toward females.

\subsubsection{Study Procedure}
We presented our participants a set of tweets collected right before the start of recruitment. 
We chose this approach because studies have shown that there is a lag in terms of the amount of time media coverage takes to influence public opinion, with some exceptions (e.g. mass shootings)~\cite{roberts2002agenda}.
As a result, we anticipated that participants would be least likely to be familiar with the most current tweets.
Participants were randomly assigned to either the treatment or the control group, with a quota check to ensure balanced allocation across political ideology.
To counter order effects, we presented tweets in random order.
\additionnew{We added attention checks--questions with options reversed--right before the Twitter feed.
Taking recommendations from Qualtrics, we also discarded participants who spent less than six minute to respond--the half of the median time spent by users in a soft-launch of 50 users.}
\addition{Participants saw each tweet and answered the questions for that item before scrolling down to the next one.
This approach reflects a natural setting of modern social media sites, where users browse feeds in real-time, click links in-situ, and the same post usually do not appear again at a later time point.}
\addition{To reduce response bias, we framed the questions to ask for credibility of the items (e.g., how do you feel about the post?) instead of the effects of nudges (e.g., how does the nudge affect your perception?). The unexpected effect on \textit{Questionable} in our formative study suggests a lack of response bias.}

\subsubsection{Method of Analysis}
We initially perform mean comparison with Mann-Whitney U-test. However, note that each user saw multiple tweets with each intervention. To model such repeated measurements for the same intervention, we further use a mixed-effects logistic regression. 
\setlength{\abovedisplayskip}{1pt}
\setlength{\belowdisplayskip}{1pt}
\begin{equation}\label{eq:1}
\mathbf{y} = X\mathbf{\beta} + Z\mathbf{u} + \epsilon
\end{equation}
In Eq. \ref{eq:1}, the response variable (credibility score) ($\mathbf{y}$) is the dependent measure for our experiment. While fixed effects ($X$) are the independent measure, random effects ($Z$) are the variables repeated in multiple observations; that is, tweets. The residual ($\epsilon$) is the error in fitting. Finally, $\beta$ and $\mathbf{u}$ are the coefficients of fixed and random effects, respectively. We used an \textbf{R} implementation of a linear mixed-effects regression model, \textit{lme4.lmer}, 
on our dataset~\citep{lme4pack56}. 

\textbf{Dependent Variable:}
Our dependent measure is the credibility score, a continuous variable computed by averaging the five credibility question responses (see Figure \ref{tb:survey}) followed by standardization. We perform robustness checks by rerunning mixed-effects ordinal logistic regressions on each of the five credibility questions. We find no significant differences in the resulting model coefficients, suggesting sufficiency in modeling the credibility score as continuous.

\textbf{Independent Variables:}
The independent variables related to RQ1 include main effects and interaction effects derived from the two experimental conditions: users' group (control or treatment) and intervention type ($T_R$, $T_Q$, $T_U$). For RQ2 and RQ3, we examine three variables, including political ideology, political cynicism, and media skepticism. For the sake of analysis, we map political ideology, measured on a seven-point Likert scale, to three groups consisting of Democrats, Independents, and Republicans. 
Following a prior scholarly work~\cite{carr2014cynics}, we used the median score across all the questions in each variable (political cynicism and media skepticism) to split the participants into two groups. In our representative US sample, the median for media skepticism was $2.75$ ($\alpha=0.72$,  $M=2.48$, $SD=0.98$) and the median for political cynicism was $4.00$ ($r=0.53$, $M=4.09$, $SD=0.81$). Similar to Carr et al., we considered values greater than the median as high on that category and vice versa~\cite{carr2014cynics}. In other words, media skepticism of 3.00 would be labeled high media skepticism, while 2.75 would be labeled low media skepticism. We also include the political leanings of the news sources used in our tweet selection procedure as an additional independent variable. 

\textbf{Control Variables:}
Prior studies indicate that the level of interest in a news story can influence users' credibility assessment~\citep{metzger2007making}. Therefore, we include participants' interest in a tweet as a control variable, measured on a five-point Likert scale ranging from low to high interest. Other control variables include users' demographics, such as gender, age, education, and Twitter usage frequency. 



\subsection{Results (Study 1)}

\begin{figure}[t]
\begin{minipage}{\textwidth}
  \begin{minipage}{0.44\textwidth}\vspace{0pt}
    \centering
    \vspace{15pt}
    \sffamily \scriptsize 
   \begin{tabular}{r|cc|l}
		& Control  & Treatment  & Bet. Subj. MWU-test \\ 
                & Avg. Cred. & Avg. Cred.& (Cohen's $d$) \\ \cline{1-4} 
 $T_R$  & 0.62 & 0.67 & 187488.0(0.162)***  \\
 $T_Q$ & 0.58  & 0.55 & 198180.5(0.072)*  \\
 $T_U$ & 0.46 & 0.37 & 171763.0(0.296)***  \\
 \cline{1-4}
 n & 693 & 597 \\
 
 \cline{1-4}
 
	\end{tabular}
    \captionof{table}{\additionnew{Mann-Whitney U test results for Study 1. Here, `n' denotes the number of tweets rated in each condition. Avg. Cred. is the mean of `n' credibility scores; * $p < 0.05$, *** $p < 0.001.$}}\label{tb:utest}
  \end{minipage} 
  \hfill
  \begin{minipage}{0.54\textwidth}\vspace{0pt}
    \centering
    \sffamily \scriptsize 
    \includegraphics[width=0.99\textwidth]{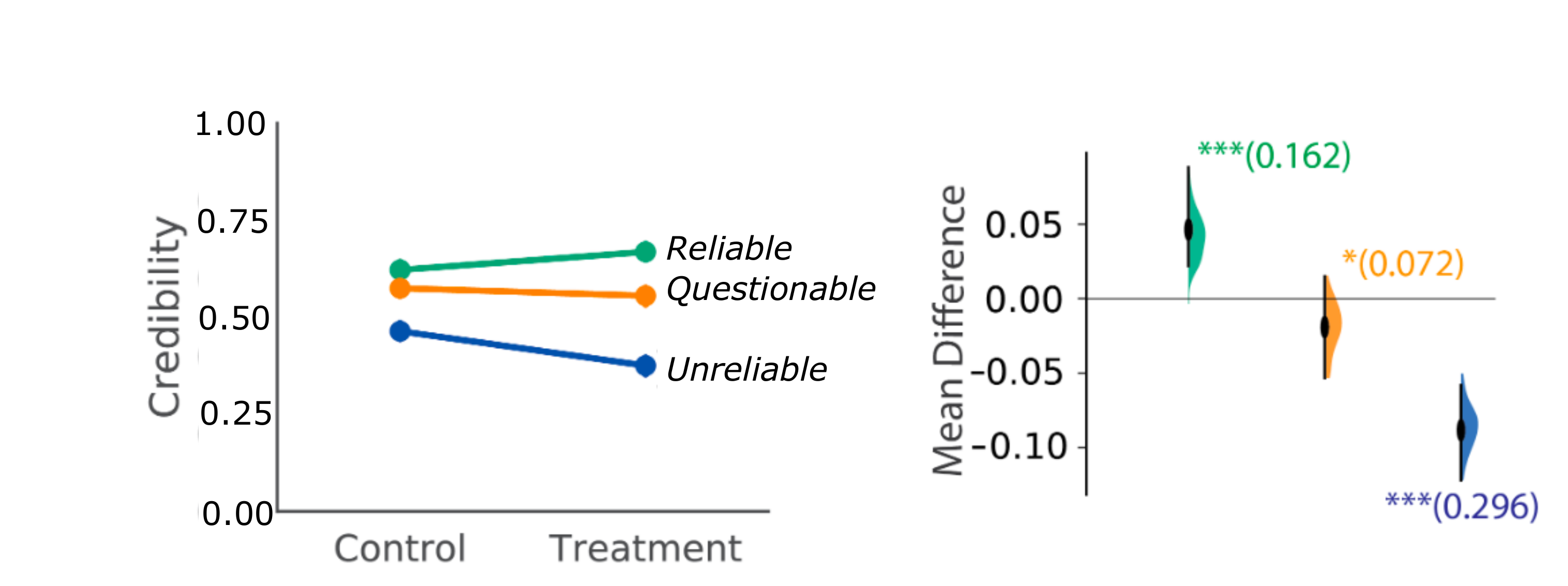}
    \vspace{-8pt}
      \captionof{figure}{Shows interaction effects between user groups and nudge types in Study 1. The numbers inside the brackets are the effect sizes, Cohen's $d$. }\label{fig:utest}
    \end{minipage}
  \end{minipage}
\vspace{-15pt}
\end{figure}

 
 
\subsubsection{RQ1: Effect of the Nudges}
For RQ1, initially we investigated our data using mean comparison. 
Table \ref{tb:utest} shows the mean values and Mann-Whitney U test results of our experiment and Figure \ref{fig:utest} shows corresponding interaction plots. 
Users in the treatment group rated the credibility of non-mainstream news tweets ($T_U$) significantly lower than did users in the control group, suggesting the effectiveness of our intervention ($Z{=}171763, p{<}0.001$, Cohen's $d{=}0.296$ 
). 
Additionally, treatment users rated mainstream news tweets without questions ($T_R$) as more credible than corresponding control users ($Z{=}187488, p{<}0.001$, Cohen's $d{=}0.162$
). 
Our participants showed significant decrease in their rating of mainstream news tweets with questions ($T_Q$) ($Z{=}198180, p{<}0.05$, Cohen's $d{=}0.072$).

\begin{table}[h]
	\centering
    { \sffamily \scriptsize 
    \resizebox{.95\textwidth}{!}{%

	\begin{tabular}{p{0.45\linewidth}p{0.06\linewidth}p{0.05\linewidth}p{0.01\linewidth}p{0.06\linewidth}p{0.05\linewidth}p{0.01\linewidth}p{0.06\linewidth}p{0.05\linewidth}}
		\toprule
		& \multicolumn{2}{c}{Base Model} & &  \multicolumn{2}{c}{Politics \& Media Model} & & \multicolumn{2}{c}{3-Way Interaction Model} \\
		\cline{2-9}
		& $\beta$      & SE  &  &  $\beta$      & SE  &  & $\beta$     & SE     \\
		\hline
		(Intercept)                           & 0.29***          &               & & 0.27***          &               & & 0.32***      &    \\
		 \textbf{Control Variables}\\
		Gender (Male)                                  & 0.03*          & 0.05          &  & 0.03*          & 0.05          &  & 0.04      & 0.07      \\
Education                             & -0.00         & -0.01         &           & -0.01         & -0.03         &           & -0.01     & -0.04             \\
Age                            & 0.00          & 0.01          &          & 0.01          & 0.03          &          & 0.01      & 0.04          \\
Social Media Usage             & 0.01          & 0.01          &       & 0.01          & 0.02          &       & 0.01      & 0.02            \\
Interest in the Tweet             & 0.09***          & 0.45          &       & 0.09***          & 0.45          &       & 0.08***      & 0.36              \\
\textbf{Experimental condition} \\
Type(Mainstream-Question)                         & -0.03         & -0.05         &         & -0.03         & -0.05         &            & -0.07      & -0.11                  \\
Type(Non-mainstream)                         & -0.16**         & -0.26         &            & -0.16**         & -0.26         &            & -0.23**     & -0.38                 \\
Group(Treatment)                                     & 0.04**          & 0.07          &    & 0.04**          & 0.07          &            & 0.08      & 0.13               \\
Type(Mainstream-Question):Group(Treatment)             & -0.06***         & -0.08         & & -0.06***         & -0.08         &           & -0.11*     & -0.14          \\
Type(Non-mainstream):group(Treatment)             & -0.10***         & -0.13         & & -0.10***         & -0.13         & & -0.21***     & -0.26                \\
\textbf{Politics and Media} \\
Ideology(Democrat)                                   & & & & 0.03*          & 0.05          &             & 0.01      & 0.01                 \\
Ideology(Republican)                                     & & && -0.03*         & -0.04         &            & -0.03      & -0.05                \\
Political Cynicism(Low)                                   & & && 0.01          & 0.02          &             & 0.01      & 0.02            \\
Media Skepticism(Low)                                   & & && 0.05***          & 0.08          & & 0.06**      & 0.11              \\
Account Leaning(Conspiracy)                                   & & && 0.07          & 0.08          &             & 0.06      & 0.07                 \\
Account Leaning(Left)                                     & & && 0.09          & 0.14          &             & 0.09      & 0.14             \\
Account Leaning(Right)                                     & & && 0.00         & 0.01         &            & 0.00      & 0.00                \\
\textbf{Experimental Condition x Other Variables}\\
Group(Treatment):Gender(Male)         & & &&               &               &                  & -0.08*     & -0.11                \\
Type(Non-mainstream):Interest in the Tweet         & & &&               &               &                  & 0.02*     & 0.11                \\
Type(Non-mainstream):Group(Treatment):Gender(Male)         & & &&               &               &                  & 0.09*     & 0.08                \\
Type(Non-mainstream):Group(Treatment):Interest In a Tweet         & & &&               &               &                  & 0.03*     & 0.10                \\

\hline
		Adj. $R^2$ (Marg./Cond.) & \multicolumn{2}{c}{ .310/.483 } & & \multicolumn{2}{c}{ .347/.494 } & & \multicolumn{2}{c}{ .358/.508 } \\
		N = 3870 & \multicolumn{8}{r}{* p\textless{}.05, ** p\textless{}.01, *** p\textless{}.001}\\
        \hline
	\end{tabular}}
	}
		\caption{Regression models on the credibility score. The base model contains nudge type, user group, control variables and the interaction between user group and nudge type. The politics and media model adds users' political ideology, media skepticism and political cynicism variables to the base model. The 3-way interaction model further includes the interactions of nudge type, user group and other variables with significant main effects in the politics and media model (Gender, Interest in the Tweet, Ideology and Media Skepticism).}
		\vspace{-25pt}
		\label{tb:regression2}
\end{table}

Our experimental setting with each user rating multiple tweets prompted us to further analyze the data using a series of mixed-effects regression models.
Table \ref{tb:regression2} shows this analysis.
To determine the effects of the experimental conditions, our base model includes news source type, group assignment, and their corresponding interactions.
We find that tweet type---mainstream or non-mainstream---is strongly correlated with a tweet's credibility score. 
Non-mainstream tweets are generally rated less credible than mainstream news sources with a small effect size ($\beta{=}-0.16$, $p{<}0.01$ 
and Cohen's $f{=}.07$). 
This result suggests that users can differentiate between mainstream and non-mainstream tweets even without our nudges. 
However, treatment users (those who received nudges) generally rated tweets as slightly more credible than control users ($\beta{=}0.04$, $p{<}0.01$, Cohen's $f{=}0.04$). 
Nevertheless, there is an interaction effect between tweet type and user group. 
Treatment users scored non-mainstream tweets lower than control users with a medium effect size ($\beta{=}-0.10$, $p{<}0.001$ and Cohen's $f{=}0.16$). 
Treatment users also rated mainstream tweets with questions as less credible than did control users ($\beta{=}-0.06$, $p{<}0.001$, Cohen's $f{=}0.16$).
The decrease in the credibility perception scores of both mainstream questioned tweets ($T_Q$) and non-mainstream ones ($T_U$) suggests that our nudges can help users consume these news items as less credible, thereby answering RQ1. 

\subsubsection{RQ2: Effect of Political Ideology}
To answer RQ2, we examine the political ideology variable in our politics and media regression model. Politically Independent users serve as the point of reference for this variable. We find that Democrats generally rated all tweets slightly higher in credibility than did Independent users ($\beta{=}0.03$, $p{<}0.05$, Cohen's $f{=}0.06$), whereas Republicans rated them slightly lower than Independents ($\beta{=}-0.03$, $p{<}0.05$, Cohen's $f{=}0.06$). 
Due to this main effect, in our 3-way interaction model, we further explore whether political ideology had any interactions with the three nudges and the users' group assignments. 
However, we find no significant interaction.
Therefore, nudges changed users' credibility perceptions irrespective of their political leanings. 
We discuss these findings later.

\subsubsection{RQ3: Effect of Political Cynicism and Media Skepticism}
To answer this RQ, we examine two variables (political cynicism, and media skepticism) in our politics and media regression model.
Between media skepticism and political cynicism, only media skepticism had a significant effect, where users with lower media skepticism rated tweets as more credible ($\beta{=}0.05$, $p{<}0.001$, Cohen's $f{=}0.10$). In our 3-way interaction model, we further explore whether media skepticism had any interactions with our key variables of interest---treatment and control groups, and the three nudges. We do not find any significant interaction effects. 
Therefore, nudges changed users' credibility perceptions irrespective of their attitudes towards politics and media. We elaborate on these findings in the Discussion section.

\subsubsection{Effect of Control Variables}
We examine whether user demographics and users' interest in a news story had any effect on how they rated the credibility of the news tweet. 
Across all three models, the effects exerted by our control variables are consistent. 
Independent of whether a user was assigned to the control or treatment group and independent of the type of news (whether mainstream or non-mainstream) they saw, users provided higher credibility scores when they were interested in a story, with a large effect (base model effects:$\beta{=}0.09, p{<}0.001$, Cohen's $f{=}0.5$). 
Among demographic variables, male users rated tweets as more credible with a small effect size (base model effects: $\beta{=}0.03, p{<}0.05$, Cohen's $f{=}0.03$). 
The remaining demographic variables did not show any significant effect~\footnote{\additionminor{Additionally, we examined whether there was any learning effect compounded from seeing multiple nudges. To do so, we added the order (from 1 to 9) of the tweets in which participants evaluated their credibility and its interaction with nudge type and user group in our regression models. We found no significant effect of the order.}}. 

\section{Study 2: Field Deployment}\label{sec-study2}
To gain insights into how we can understand and improve the current design nudges for credibility, we conducted a qualitative study. 
For this purpose, we recruited participants to use \FeedReflect{} on Twitter for five days.
This process allowed us to evaluate it in a more ecologically valid environment than Study 1~\cite{brewer2000research}.
Below, we describe the process.
%

\subsection{Method (Study 2)}
\subsubsection{Recruitment}
To recruit users for this study, we used Twitter advertising, following a prior strategy~\cite{im2020synthesized}. 
We devised several targeting mechanisms to promote the advertisement to our desired group, including age range (>=18), language (English), location (USA) and whether users followed top mainstream and non-mainstream news accounts in our list. 
Initially, we were not successful in getting responses from broader regions, so we iteratively revised the location to target nearby states for effective recruitment.
Additionally, we promoted the advertisement within our academic network on Twitter. 
From 50 interested participants, we recruited 12 participants for the study by filtering our spams and users with less than 100 followers . 
Overall, our participant group consisted of 5 females and 7 males with an average age of 29.5 (std. dev. = 6.5) and a political tilt towards Democrats (Democrat = 6, Independent = 4, Republican = 2).

\subsubsection{Procedure}
Followed by an informed consent process, we instructed users to install the \FeedReflect{} browser extension.
To promote a natural setting, we encouraged users to use Twitter during the five-day study as they normally would.
However, it is possible that Twitter's news feed algorithm may not have surfaced news items on their feed each time. 
Hence, we also encouraged users to visit the Twitter profiles of some news sources each day to ensure that users experience how \FeedReflect{} works. 
After five days, we asked them to fill out a post-study survey on demographic details followed by an interview in a semi-structured manner (see Appendix \ref{int:ques} for the interview questions). 
To facilitate their responses, we asked users to walk us through their news feed during the interview.
Each participant received a \$30 gift card for participating.

\subsection{Results (Study 2)}
We analyzed the interview data using a grounded theory approach~\cite{strauss1994grounded}. 
The first author transcribed the audio from the interviews and analyzed the data to come up with a set of themes. 
These themes were discussed and refined with the other authors. 
Below, we present our final set of themes.

\subsubsection{\FeedReflect{} facilitates more conscious news consumption} 
%

Most of our participants (9/12) provided positive feedback on their overall experience with \FeedReflect{}, referring to the design and application.
Some participants (U1,U6,U9) particularly mentioned that they liked the bandwagon heuristic with the questions in replies. 

\customquote{It [\FeedReflect{}] quickly highlights. So you know what to look for. Especially when it's a question mark, I do actually open up the comment section and read what the question is.}{U9}

Others liked it because it served as an educational tool to ``train the user to thoughtfully think about'' news (U4) or because it did not add overhead to their current Twitter experience (U2). 
Overall, users reported two phenomena. We describe them below.

\noindent \textbf{Improved Recognition of News Content and News Genres:} 
One of the impacts that participants (5/12) mentioned was the perceived difference in the amount of news content in their feed compared to their prior experience. 
For example, they perceived that there was more news content in their feed than before.
\FeedReflect{} even helped some participants pay more attention to the types of content that news sources produced. 

\customquote{It [\FeedReflect{}] really just told me that NPR produces more articles that are opinionated [the participant was referring to an Op-Ed] on Twitter than I thought.}{U1}

\noindent The article labeled as \textit{Questionable} made them realize that it was an op-ed article, rather than news. 

\noindent\textbf{Attention towards Bandwagon Heuristic:} 
Out of the three nudges, participants (7/12) noticed the \textit{Questionable} tweets highlighted in yellow the most, and the number of questions below them.

\customquote{I noticed the one [question icon] at the bottom more ... most people who use Twitter a lot ... 9 out of 10 are more likely in tune with the replies/retweets. Usually I use those as a sign of popularity.}{U1}

\customquote{If I see it [a tweet] as yellow ... I do get the information that this is either ... a lot of people don't like the news article or this article might have controversial or incorrect facts.}{U10}
While users, as U1 indicated, may see traditional retweet or reply numbers as an indicator of popularity, one participant (U10) correctly pointed out the bandwagon cue as an indicator of controversy.
Thus, nudges imitating existing designs on social media can be useful.

Overall, these phenomena support that our nudges can improve users' news perception in two ways: (i) with an overall impression of total news items on users' feeds broken down based on the reliability of sources, facilitating better perception on its genres; and (ii) with individual attention towards particular news items.



\subsubsection{Concerns in Using Heuristics for Nudging News Credibility} 
Interviews also revealed two concerns regarding our nudge design. 
We discuss these concerns below.

\noindent \textbf{Trust Issues with Heuristics:}
A majority of our participants (7/12) questioned the use of bandwagon heuristic to differentiate \emph{Reliable} and \emph{Questionable} news items.
Because audience composition can vary by the source and the topic of a news item, and influence bandwagon heuristic, they were concerned about its disparate impact.
One participant pointed out that followers of the New York Times (NYT) are comparatively more diverse than followers of Fox News. Consequently, audiences with an opposing stance on a news report from the NYT may question it. In contrast, Fox News, having a homogeneous audience which mostly supports its reporting, may hardly question their reports. Therefore, our bandwagon heuristic would be skewed based on the audience of a report.
%
\customquote{NYT [The New York times] and MSM [mainstream media] in general have a lot more reactions from skeptical readers given the current administration. And to some, the color-coding and the number of questions may indicate that the news is subjective or ``fake'' when you compare it with other outlets such as Fox News that have fewer reactions on Twitter and a more homogeneous audience.}{U7}


These responses suggest that even though a user may understand the bandwagon heuristic, the heuristic itself may have some shortcomings, which makes it challenging for the user to trust it as a metric for gauging credibility. 



\noindent \textbf{Adverse Effects of Nudges:}
Our participants (2/12) suggested two adverse effects of the nudges. One participant (U11) proposed that users may use the bandwagon heuristic based on like-minded questions as a justification for their attitude-consistent belief in politically opposing news, thus promoting a \textbf{confirmation bias}.

\customquote{If I agree with you and you are questioning an article that I questioned as well ... since you personally agree with me, it confirms my bias against that piece of information.}{U11}

A similar effect has been suggested by scholars in the past, who show that rating mechanisms on online platforms (e.g., Facebook ``likes'') may guide users' news selection process~\cite{messing2014selective}.
If such effects exist, users may become more polarized. 
Compared to confirmation bias stemming from the existence of a nudge, the \textbf{absence of nudges}, mentioned by U4, could also have a harmful effect.

\customquote{I think the ones that I subconsciously ignore are the ones that have no color at all. If there aren't any flags ... no color blocks, I am more inclined to assume that the content is valid.}{U4}

This participant suggested that the absence of nudges creates an illusion of validity of content without nudges. 
Indeed, recent research points out the same phenomenon when false reports are not tagged, resulting in a false sense of being validated~\cite{pennycook2020implied}.
One way to address this concern is, again, to be transparent about not being nudged with an additional tool tip message for the news items that are not nudged.

Overall, our participants' concerns suggest that designers need to evaluate two aspects of nudges: (i) How trustworthy the design components of the nudges are (ii) Whether the presence and absence of nudges adversely affect users.



\subsubsection{Opportunities for Credibility Nudge Design}
In addition to differentiating news credibility, we asked participants what other functions they would like in \FeedReflect{}. 
Participants suggested improvements in three directions.

\noindent \textbf{Extending the News Source List:} Our participants were concerned on the limited set of news sources we considered. They (5/12) suggested that they often see misinformation from non-news entities, including their acquaintances. To allow more news identification, some (2/12) asked us to include local news sources. 
With our participants following diverse sources for information, our limited news source list was naturally inadequate for them.


\noindent \textbf{Indicating News Genres and Reporting Biases:} Suggestions from the participants included distinguishing opinion items from news (3/12) and indicating bias in a report as a nudge (2/12). 

\customquote{Give a notification that say what I am seeing is an op-ed rather than straight facts.}{U4}
\customquote{Is it possible to state that this news article is biased towards this particular claim?}{U10}

A recent survey shows that about 50\%  US adults ``are not sure what an op-ed is'' and that about 42\% of respondents perceive that opinion and commentary are often posed as news in most news articles~\cite{American49online}. Therefore, a significant share of the population may appreciate having nudges that differentiate op-eds from news as well as other indicators of bias stems. Incorporating such attributes in nudging might help users better determine the credibility of news content.

\noindent \textbf{Curbing Misinformation Sharing:} To prevent the sharing of misinformation, some participants (2/12) proposed implementing nudges on share (or retweet) buttons.

\customquote{[when someone clicks the share button] If there is a notification that says this source is not credible then people would be less likely to share it.}{U2}

Research indicates that about 59\% of links shared on Twitter have never been clicked~\cite{gabielkov2016social}, i.e., users often share news items without reading them.
If nudges can help user determine the unreliability of news from misinformative sources, they might be less likely to share them.


In summary, our participants proposed improvements to our nudge design in three key areas: (i) improving existing classifications by extending the source list, (ii) expanding news classifications of nudges in alternate areas, and (iii) targeting users' interactions with news on social media.

\section{Discussion}
Below, we elaborate on our results, starting with each research question from Study 1, followed by opportunities and challenges suggested by the participants in Study 2.

\subsection{RQ1: Effect of Nudges on Credibility}
Our regression analyses in Study 1 revealed that users' credibility ratings were considerably different between the treatment (nudged) group and the control group.
While other nudge designs have proven effective in reducing sharing intentions of misinformative content, their effectiveness was shown on a particular news genre, such as COVID-19 and HIV~\cite{pennycook2020fighting,nekmat2020nudge}. 
In contrast, we examined the effects of our nudges on a wide variety of popular news items surfacing over multiple days, thus offering a more generalized result.
Our intervention provides a less authoritative approach that gives users simple but transparent information for them to make their own judgments. 
News feeds, as they typically present limited information on social media, have few features to distinguish content quality. 
Tweets from bots, friends, mainstream news, and non-mainstream sources are all given equal weight in terms of visual representation in any feed, making it difficult for users to sift through them.
Though people are capable of identifying misinformation, social media makes it challenging to make informed analytical judgments~\cite{pennycook2019understanding}.
Our results suggest that users might appreciate it if social media sites provide tangible signals to work through this clutter, which is further exemplified by participants' suggestion to differentiate news and op-eds in Study 2.

Apart from facilitating better perceptions of the credibility of news, \FeedReflect{} may also act as a ``translucent system''~\cite{erickson2000social}. The theory of social translucence posits that we should aim for systems that make online social behavior \emph{visible} to facilitate \emph{awareness} and \emph{accountability}. Note that our participants in Study 2 suggested improved recognition of particular types of news content on their feed and were more aware of what they were seeing on their feeds. Our nudges on news content that are liked or shared by users' followers or friends could also have similar impacts, wherein users become more \emph{aware} of their peers' news consumption behaviors. 
When their peers like/share misleading news, one may hold the peers \emph{accountable} by pointing out the misleading content. 
Besides, after seeing the nudges on unreliable content, users may restrain themselves from sharing such content and reflect on their sharing habits.


\subsection{RQ2: Influence of Political Partisanship on Nudge Effects}
Our regression results suggest that \FeedReflect{} changed users' perception of credibility irrespective of their political views; that is, there were no interaction effects between political characteristics and the effects of interventions.
\addition{This result is consistent with recent studies showing the success of interventions in limiting sharing intentions of misinformation, irrespective of users' political affiliation~\citep{wood2019elusive,roozenbeek2019fake}.}
Although some prior literature argue that citizens may view nudges as partisan policies and may not support nudges when they conflict with users' partisan preference~\cite{sunstein2016people}, other scholars suggest that this behavior can be countered by reducing partisan cues in nudges~\cite{tannenbaum2017misplaced}.
We incorporated this suggestion in our design by showing nudges on news content from all political leanings and nudging news content in both directions (Reliable, Questionable, and Unreliable). 
However, in practice, users tend to typically follow news based on their partisan preferences~\cite{kohut2012cable}.
In such a setting, users who follow only alternative fringe sources may see mostly \emph{Unreliable} nudges triggered on their reports and perceive \FeedReflect{} as partisan.
One potential design solution is to show the similar news item from reliable sources, with the same partisan view, to help them understand the alternatives.


\subsection{RQ3: Influence of Political Cynicism and Media Skepticism on Nudge Effects}
In our study, we did not find any impact of political cynicism or media skepticism on nudge effects. 
This convincing nature of our nudges---that nudges worked irrespective of users' prior media skepticism and political cynicism---is promising.
\addition{Our result for media skepticism is aligned with a recent work where media skepticism did not affect nudge effects~\cite{nekmat2020nudge}.} 
Research suggests that media skeptics, despite significant exposure to alternate sources, still seem to have moderate to high consumption of mainstream news~\cite{tsfati2005people}. 
Therefore, our nudges could improve news credibility assessment of both mainstream and non-mainstream sources by skeptics in the wild.
Scholars suggest that exposure to fake news mediated by belief in its realism increases political cynicism~\cite{balmas2014fake}.
Thus, if nudges can reduce belief in fake news, it could help mitigate increasing cynicism towards politics.
Furthermore, our nudges can be utilized as an alternative to censorship by social media, thus helping mitigate the concern that social media apply censorship in a disparate manner across different political affiliation, as raised by participants in our formative study.


\subsection{Opportunities in Designing News Credibility Nudges}
Our field deployment of \FeedReflect{} showed several opportunities in designing credibility nudge in the future. 
First, participants' attention to the bandwagon heuristic reveals how designers can utilize existing Twitter infrastructure in their design.
Though the impact of the bandwagon effect in collaborative filtering has been discussed in the literature~\cite{sundar2008bandwagon}, it has been underutilized in a news credibility context. 
Our study suggests that applications similar to ours can act as valuable markers of information credibility. 
Second, participants seeking nudges on a wider set of sources (e.g., news and non-news sources), and alternate types of taxonomies (e.g., news and op-eds) suggests their need for tools to differentiate information reliability.
Comparatively, nudges on tweet actions (e.g., retweet and share) may play a stronger role in curbing the spread of misinformation, as research indicates that sharing without thinking is often weaponized for this purpose~\cite{Misinfor57online}.
For example, when users click on the share button, activating a ``timer nudge''---a visual countdown to complete the action---could make users rethink before they share~\cite{wang2013privacy}.

\subsection{Challenges in Designing Nudges with Heuristics}
Our final evaluation presented several challenges in designing credibility nudges. 
First, participants showed their skepticism towards the selection of heuristics (e.g., bandwagon heuristic) in design.
Though the bandwagon heuristic can be robust and hard to falsify, in an open forum such as Twitter, it is open for manipulation. 
Perhaps, as one of the participants suggested, feedback on the validity of the question may be helpful. 
Still, problems may also exist with feedback. 
For one, due to partisan attitude, feedback wars, similar to edit wars in Wikipedia, might result~\cite{sumi2011edit}.
Additionally, due to frequent updates and the scale of social media content, feedback from volunteers might be scarce on most items.
Perhaps a system designer can show the distribution of feedback by users' leanings to reduce the effects of feedback wars and solicit user contributions by promoting items with scarce feedback.
Second, participants' concerns with the bandwagon heuristic promoting confirmation bias might be an extension of prior findings that users tend to prefer selecting information (including political information) consistent with their preexisting beliefs~\cite{knobloch2009looking}. 
However, scholars have shown that the extent of confirmation bias in information selection, particularly in news selection, is small~\cite{hart2009feeling}.
In the event of confirmation bias, we can computationally de-bias the heuristic cue, as in prior works~\cite{adomavicius2014biasing}.
Lastly, audience misperceptions of non-nudged content being credible indicate an additional challenge in design. 
This effect has also been demonstrated in a recent work~\cite{pennycook2020implied}.
One way to solve this problem would be to add nudges to all content. 
Aside from these challenges, one participant (U7) pointed out that \textit{``News might change and the user will not see the update when more legitimate questions are added to the replies''}.
As user interactions accumulate over time, the number of questions in replies could change, wherein the same tweet would be categorized as \emph{Reliable} at first and \emph{Questionable} at a later time. 
This change in nudge category stemming from our choice of the bandwagon heuristic could imply an inconsistency in nudge design.
System designers can incorporate delayed intervention mechanisms and inform users of this cold-start issue. 
Overall, these challenges inform designers  about considerations for designing nudges with heuristics.

\section{Implications And Opportunities for Designing 
Credibility Nudges} \label{sec:design_impl}
\begin{figure}[t]
\centering
\vspace{-15pt}
\includegraphics[width=0.9\textwidth]{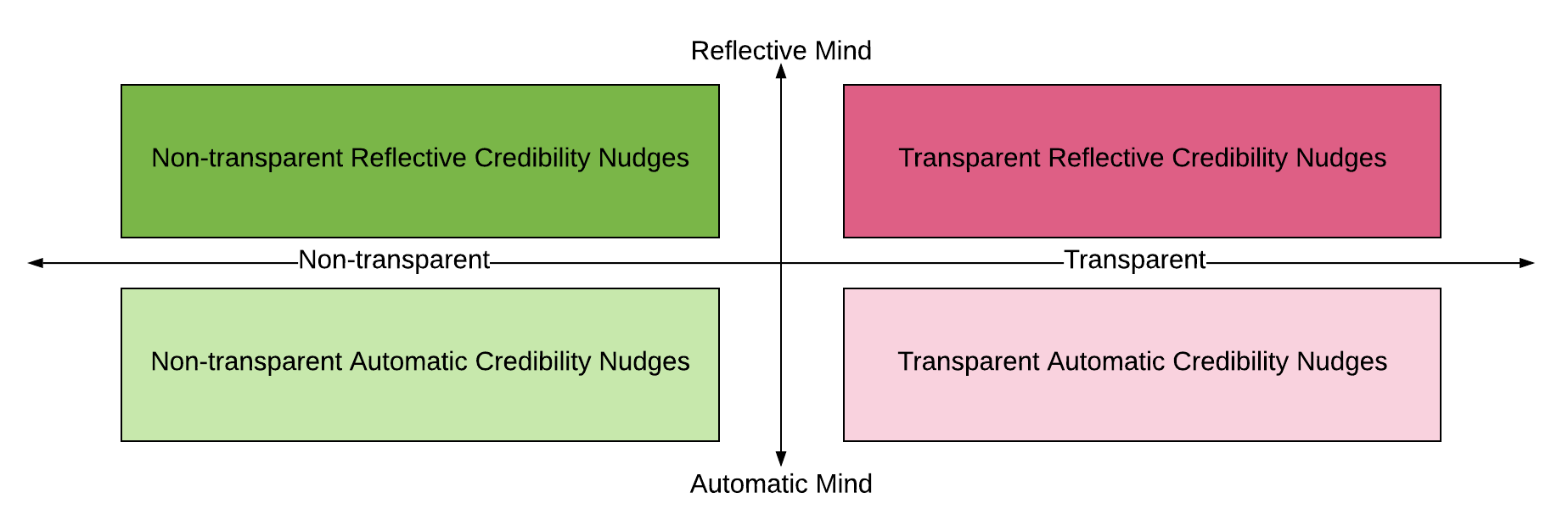}
\vspace{-17pt}
\caption{Types of nudges based on transparency and mode of thinking. This figure emulates Figure 1 by Caraban et al.~\cite{caraban201923}. This work lies in the bottom-right quadrant.} \label{fig:nudge_type}
\vspace{-16pt}
\end{figure}
We built \FeedReflect{} by fusing the theoretical background on nudges with theories underpinning credibility evaluation. Researchers have the opportunity to explore the nudge categories we built our design around, experiment effectiveness of other heuristics and utilize alternate nudging method. Below, we elaborate a few possibilities around these areas while discussing potential considerations.

\paragraph{Exploring Additional Nudge Categories}
Referring to the four categories of nudges, divided along the two dimensions of \emph{transparency} and \emph{mode of thinking} (automatic or reflective), we have illustrated how \FeedReflect{} resides on one of the four quadrants---\emph{transparent automatic} (see Figure \ref{fig:nudge_type}).
\addition{Nudge theorists describe transparent nudges with reflective and automatic mode as, respectively, reflective decision making and instinctive decision making~\cite{hansen2013nudge}.
Technology-mediated nudging research focus more on transparent reflective quadrant~\cite{caraban201923}.
For example, Facebook's ``Related Articles'' feature to tackle misinformation exemplifies such a nudge design~\cite{Designin7online}.
Twitter's blue check ``verified'' marker on profiles is another example of a transparent automatic credibility nudge.
\additionnew{Between reflective and automatic mode of thinking, each has their own benefit. For example, while nudges with automatic mode of thinking can be diminished over time, reflective mode of thinking can educate users and have a lasting impact.
On the other hand, design considering reflective mode of thinking requires additional considerations such as motivating the users and assisting in reflection~\cite{adams2015mindless}.
For example, to motivate users we can show statistics of how often users misread a news tweet as a nudge and to assist them in reflection this nudge can include the list of the most common mistakes.
}
For non-transparent quadrants, scholars propose reflective ones as a manipulation of behavior and automatic ones as a manipulation of choice, in both cases without users' awareness.
\additionnew{
An example of non-transparent automatic nudging could be showing the average reading time of an article which could prompt users to think shorter articles as less detailed and less credible.
Comparatively, research show fewer work in non-transparent reflective quadrant~\cite{caraban201923}. 
Due to their deceptive nature, designers need to consider ethical consideration while experimenting on non-transparent quadrants.
}
Overall, this is the start of work in this space; much research needs to be done in cataloging and evaluating news credibility nudges along the other dimensions spanning the four quadrants. 
}

\paragraph{\addition{Exploring Alternate Heuristics for Nudging Credibility}}
Our nudge design is based on two heuristics under the \textbf{A}gency affordance, drawn from the MAIN model's list of technology-mediated affordances affecting credibility judgment.
Designers have the opportunity to explore the other heuristics in their design. For example, designers could offer design cues to distinguish news items with/without video footage and prompt a \textit{realism heuristic} from \textbf{M}odality affordance.
Or, design cues distinguishing news tweets with interactive content (e.g., interactive 3D footage or charts) could prompt \textit{interaction heuristic} from \textbf{I}nteractivity affordance. For \textbf{N}avigability affordance, designers can prompt \textit{browsing heuristic} by providing additional hyperlinks to journalists' past activities (e.g., MuckRack profile~\cite{MuckRack33online}) besides news items.

\paragraph{\addition{Examining Alternate Nudging Method}}
While the original proposers of the concept did not lay out a fixed method for creating successful nudges~\cite{selinger2011there},
Caraban et. al. recently devised six broad categories for 23 nudging methods used in prior HCI works, namely, \texttt{Facilitate} (e.g., \textit{default choice} and \textit{hiding}), \texttt{Confront} (e.g., \textit{remind consequence} and \textit{provide multiple viewpoint}), \texttt{Deceive} (e.g., \textit{add inferior alternative} and \textit{deceptive visualization}), \texttt{Social Influence} (e.g., \textit{enable comparison} and \textit{public commitment}), \texttt{Fear} (e.g., \textit{reduce distance} and \textit{scarcity}) and \texttt{Reinforce} (e.g., \textit{ambient feedback} and \textit{subliminal priming}) \cite{caraban201923}. Under these categories, \FeedReflect{} utilizes two methods from two categories.  First, it works as a \texttt{Facilitate} category by facilitating decision making through a combination of color-coding and translucent \textit{hiding} method. Second, it operates as a \texttt{social influence nudge} category with \textit{enabling social comparison} method through the use of number of questions asked by other responders. Technology-mediated credibility nudges can utilize other methods from this classification. For example, similar to NewsCube~\cite{park2009newscube}, designers can use \texttt{confront} category by \textit{offering multiple viewpoint} method on news items.
Or, flashing keywords around news items (e.g., ``reliable'' or ``questionable'') utilizing \textit{subliminal priming} method under \texttt{Reinforce} category~\cite{pinder2015subliminal}, could affect users' credibility perception of those items.

\paragraph{\addition{Designing Against Adversarial Agents.}}
Our approach to be transparent suggests that adversarial entities, upon knowing algorithmic detail, can manipulate the system. 
For example, by commenting on a factual mainstream news with a question, they can create a misperception of it being questionable.
This is a problem that most online platforms struggle with--balancing between being transparent about their algorithm while safeguarding against adversaries. Indeed, platforms like Reddit while publishing their ranking algorithms include ``some fuzzing'' to avoid manipulation of their voting system \cite{Redditov4online}.
Hence, some opaqueness in the algorithm might be desirable.
\additionnew{At the same time, platform developers could also misuse this fuzziness in the algorithm for their own benefit, such as to drive engagement~\cite{HegotFac87online}.
Indeed, Twitter's content moderation practice has been controversial in the past, in some cases resulting in reversal of moderation decisions~\cite{Initslat46online}.
Similar controversy could arise regarding nudging policy.
Therefore, designing nudges requires consideration for the multiple stakeholders of a social platform---platform developer, consumers and news producers. 
Designers would need to consider the degree to which a nudge would be resistant to adversarial attacks by each stakeholder. 
For example, crowds' question based bandwagon heuristic has a high-level of susceptibility of manipulation by the consumers compared to the platform developers and news producers. On the other hand, our authority heuristic is more susceptible to manipulation by the platform designers compared to the news producers and consumers.
Overall, a potential solution to this problem would be creating a collaborative standard authorized by all stakeholders.
For example, Facebook has already created a third-party oversight board for content moderation~\cite{Oversigh91online}.
A similar strategy can be applied to determine nudging criteria.
}

\paragraph{Considering Shortcomings of Nudging as a Design Approach}
Despite its success, nudging has its own shortcomings.
Prior literature proposes that nudges may have diminished effects in the long term for two reasons: (i) nudges relying on automatic cognition may fail to educate the user, and (ii) prolonged exposure may have various effects such as transform nudge effects into background noise~\cite{lee2011mining,sunstein2017nudges}, invoke a feeling of intrusiveness~\cite{wang2014field}, and reduce users' perception of autonomy~\cite{lehmann2016changing}.
Such effects may lead to unforeseen consequences. 
For example, a default choice nudge promoting vaccination that reduced users' perception of autonomy resulted in users unsubscribing from the program~\cite{lehmann2016changing}.
In our case, if users repeatedly encounter their favored political news source labeled as \emph{Questionable} or \emph{Unreliable}, they could become averse to the design.
\additionnew{However, designers can apply several strategies to counter this problem. On one end, they can alter the design over time or prompt users to change the intervention settings over time~\cite{kovacs2021not}. 
As an alternate, they can also choose to deploy reflective nudges which are less susceptible to the diminishing effect. A potential problem with the altering design is that news consumers may need to re-learn how nudges operate.
Regardless, designers would first have to understand the rate at which nudge effects diminishes, a direction for future research.
}

\section{Limitations and Future Work}
Our work is not without  limitations.
\addition{Our Study 1 was conducted in a controlled setting as opposed to in the wild. 
However, we see two reasons that suggest that our results demonstrating the utility of nudges in credibility assessment could extend to naturalistic Twitter setup as well. First, research shows that self-reported measures pertaining to social media usage correlate with observed behavior in the wild~\cite{guess2019accurate,pennycook2020implied}. Second, large-scale survey-based research on nudges pertaining to news on social media, show that nudges affect related attitude, such as sharing intention of misinformation~\cite{nekmat2020nudge,pennycook2020fighting}.}
Our second limitation relates to choice of population. Because we tested variables (e.g., political ideology and media skepticism) that has different meaning across countries, we had to limit our experimental population to the US. To generalize our findings to the global population, future research could replicate our study in the context of each country.
Third, our recruitment had limitations that are characteristic of any online service-based recruitment.
Though we may not have obtained the true nationally representative sample in the US, research suggests that Qualtrics provides reasonably representative US sample (approximately 6\% deviance from the US census) around demography such as gender, age, race and political affiliation~\cite{heen2014comparison}.
\addition{Overall, a large-scale Twitter deployment might reconcile these concerns in the future. 
We initially attempted to do so by contacting a large number of Twitter users, without much success due to a lack of platform-level cooperation (Our research account was repeatedly blocked). While a 2015 study had successfully piggybacked on Twitter's infrastructure to run large-scale recruitment efforts on the platform \cite{grevet2015piggyback}, we were unable to do so, despite following similar strategy. We anticipate that changes to Twitter's policies may have prevented us from running large-scale recruitment on the platform \cite{TheTwitt42online}.  }

\section{Conclusion}
In this study, we provide evidence that a nudge-based design that directs users' attention to specific social cues on the news can affect their credibility judgments. We used three nudges: \emph{Reliable}, applied to mainstream news tweets without questions in replies; \emph{Questionable}, applied to mainstream news tweets with questions in replies; and \emph{Unreliable}, applied to non-mainstream news tweets. Our experiment suggests that users who saw tweets with \emph{Reliable} nudge as more credible, and tweets with \emph{Questionable} and \emph{Unreliable} nudges as less credible compared to the control users. Moreover, our nudges were not affected by users' political preferences, political cynicism, and media skepticism. Through interviews, we found evidence of how nudges can impact users' news consumption and how the current design can be improved. This research proposes further exploration of nudge-based system design approaches for online platforms. 

\section{Acknowledgements}
\additionminor{This paper would not be possible without the  support from National Science Foundation through grant \#$2041068$. We also appreciate valuable feedback from the members of the Social Computing Lab at Virginia Tech and University of Washington.}

\bibliographystyle{ACM-Reference-Format}
\bibliography{sample-base}
\received{January 2021}
\received[revised]{April 2021}
\received[accepted]{July 2021}

\appendix
\newpage
\section{Example Tweets Used in Study 1}\label{sec-example}
\begin{figure}
    \centering
    \includegraphics[width=0.95\textwidth]{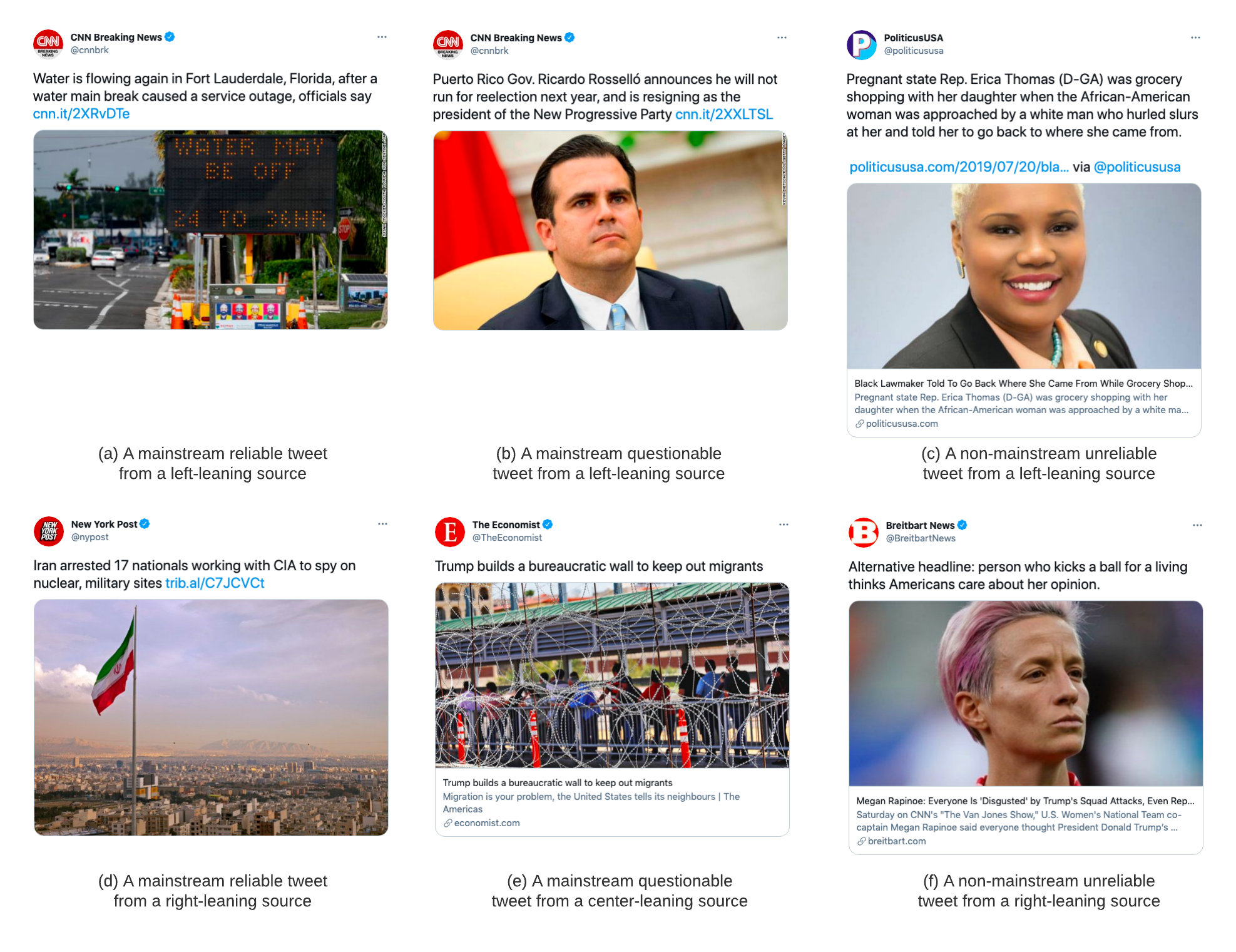}
    \caption{\additionnew{Sample tweets used in Study 1 without the interventions. The examples include reliable, questionable and unreliable tweets from left-/center-/right-leaning sources. Here, there is a mix of politically contentious (e.g., immigration, racism and LGBTQ+) and not so contentious issues (e.g., flood and national security).}}
    \label{fig:tweets}
\end{figure}

\additionnew{Figure \ref{fig:tweets} shows several tweets used in Study 1. These tweets were selected by finding the most popular tweets from the last 48 hours from various partisan sources. Notice the partisan nature of the items including immigration, racism and LGBTQ+ issues.}

\section{Study 2: Interview Questionnaire}\label{int:ques}

\begin{itemize}
    \item Could you tell me about your news reading on twitter? How often do you read news? and what type of news do you read?
    
    \item How often do you come across misinformation?
    \item Have you ever felt the need for any tools to improve news reading on twitter?
    \item Have you ever used any tools to improve news reading on twitter? What tool? How did it work?
    
    \item (Asking them to share their screen for twitter feed) How would you compare your Twitter use during the study to how you normally do? 
    \item Could you tell me about a time when you paid more attention to a news on Twitter in the last 5 days? Why? What was it about?
    \item How satisfied were you with what you saw on Twitter in the last 5 days? 
    
    \item What aspect of the intervention did you notice most? How did that impact you?
    \item When using the extension, did you think of anything it was missing? What more should it do?
    \item Can you think of any other application of this extension that you would like? 
    
    \item During the study, did you have any issues with \FeedReflect{}? Is there any aspect of the usability (eg., design, speed, accuracy) you thought could be improved?
    \item How did you feel about the plugin overall? What did you like about it? What did you dislike about it? 
    \item Would you continue using it after this study?
\end{itemize}

\end{document}